\documentstyle[epsf,times]{mn2e}

\newcommand{\simgt}{\lower.5ex\hbox{$\; \buildrel > \over \sim \;$}}
\newcommand{\simlt}{\lower.5ex\hbox{$\; \buildrel < \over \sim \;$}}
\newcommand{\bm}[1]{\mbox{{\it \boldmath$#1$}}}
\newcommand{\kaco}[1]{\left\langle{#1}\right\rangle}
\newcommand{\skaco}[1]{\langle{#1}\rangle}

\newcommand{\apj}{ApJ}
\newcommand{\prd}{Phys.~ Rev.~ D.}
\newcommand{\prl}{Phys.~ Rev.~ Lett.}
\newcommand{\mnras}{MNRAS}

\newcommand{\colskip}{@{\hspace{0.3in}}}
\newcommand{\baredth}{\;\overline{\raise1.0pt\hbox{$'$}\hskip-6pt
\partial}\;}
\newcommand{\edth}{\;\raise1.0pt\hbox{$'$}\hskip-6pt\partial\;}

\begin{document}
\onecolumn \title[Lensing tomography]{
Cosmological parameters from lensing power spectrum and bispectrum tomography}

\author[M. Takada \& B. Jain]
{
Masahiro Takada\thanks{E-mail: mtakada@hep.upenn.edu}
and Bhuvnesh Jain\thanks{E-mail: bjain@physics.upenn.edu} \\
Department of Physics
and Astronomy, University of Pennsylvania, 
Philadelphia, PA 19104, USA
}

\pagerange{\pageref{firstpage}--\pageref{lastpage}}

\maketitle

\label{firstpage}
\begin{abstract}
We examine how lensing tomography with the bispectrum and power spectrum
can constrain cosmological parameters and the equation of state of dark
energy. Our analysis uses the full information at the two- and
three-point level from angular scales of a few degrees to 5 arcminutes
($50\le l \le 3000$), which will be probed by lensing surveys.  We use
all triangle configurations, cross-power spectra and bispectra
constructed from up to three redshift bins with photometric redshifts,
and all relevant covariances in our analysis.

We find that the parameter constraints from bispectrum 
tomography are comparable to those from power spectrum tomography. 
Combining the two improves parameter accuracies by a factor of $3$
due to their complementarity.
For the dark energy parameterization $w(a)=w_0+w_a(1-a)$, the marginalized
errors from lensing alone are $\sigma(w_0)\sim
0.03f_{\rm sky}^{-1/2}$ and $\sigma(w_a)\sim 0.1 f_{\rm sky}^{-1/2}$. 
We show that these constraints can be further 
improved when combined with measurements of 
the cosmic microwave background or Type Ia supernovae. 
The amplitude and shape of the mass power
spectrum are also shown to be precisely constrained. 
We use hyper-extended perturbation theory to compute
the nonlinear lensing bispectrum for dark energy models. 
Accurate model predictions of the bispectrum in the moderately
nonlinear regime, calibrated with numerical simulations, 
will be needed to realize the parameter accuracy we have 
estimated.  Finally, we 
estimate how well the lensing bispectrum can constrain a model with 
primordial non-Gaussianity. 
\end{abstract}
\begin{keywords}
 cosmology: theory --- dark energy ---  gravitational lensing --- 
large-scale structure of universe
\end{keywords}

\section{Introduction}

Various cosmological probes have given strong evidence that a
dark energy component, such as the cosmological constant, constitutes
approximately $70\%$ of the total energy density of the universe. 
The most striking evidence comes from observations of supernovae in distant
galaxies (Riess et al. 1998; Perlmutter et al. 1999, 2000), mapping
temperature anisotropies in the cosmic microwave background (CMB) sky
(e.g., Spergel et al. 2003) and detections of the integrated
Sachs-Wolfe effect via the cross-correlation between the CMB and the
large-scale structure (Boughn \& Crittenden 2003; Nolta et al. 2003; 
Scranton et al. 2003).
If the dark energy is constant in time, its natural 
value is 50 to 120 orders of magnitudes larger than the
observed value (Weinberg 1989).  A model that allows the dark energy 
to dynamically evolve could avoid this fine tuning problem
(see Ratra \& Peebles 2003 for a review).  
Such a dark energy model can be characterized by a 
time-dependent equation of state
$w(a)=p_{\rm de}/\rho_{\rm de}$ with $w\le 0$ (the cosmological constant
corresponds to $w=-1$). It is useful to treat this as a
parameterization to be determined empirically, 
due to the lack of compelling models for the dark energy 
(e.g. Turner \& White 1997).  

The dynamically evolving dark energy affects the expansion rate of the
universe, which in turn alters the redshift evolution of mass
clustering.  It is well established that weak
lensing can directly map the mass distribution along the
line of sight by measuring the correlated distortion of images of distant
galaxies (see Mellier 1999; Bartelmann \& Schneider 2001 for 
review; and also see, e.g.,  Hamana et al. 2003, Jarvis et al. 2003, Van
Waerbeke \& Mellier 2003 and references therein for the state of
observations).  Future wide-field multi-color surveys 
are expected to have photometric redshift information for distant
galaxies beyond $z=1$ (see e.g. Massey et al. 2003). 
This additional information is extremely valuable in that it
allows us to recover radial information on the lensing field, 
which probes the redshift evolution of the expansion
history and mass clustering.  Since the weak lensing observables
are predictable {\em ab initio} given a cosmological model, lensing
tomography is a well-grounded means of constraining dark
energy evolution (Hu 1999, 2002a,b; Huterer 2002; Futamase \& Yoshida 2001; 
Abazajian \& Dodelson 2003;
Heavens 2003; Refregier et al. 2003; Benabed \& Van Waerbeke 2003; Jain
\& Taylor 2003; Bernstein \& Jain 2003; Simon et al. 2003).
While luminosity distance measures of supernovae are an established 
direct probe of dark energy
(e.g., Chiba \& Nakamura
2000; Huterer \& Turner 2001; Tegmark 2002; Frieman et al. 2003; Linder 2003),
it is important to perform cross checks of various
methods to understand systematics and because they have sensitivities
to different redshift ranges.

Non-linear gravitational clustering induces
non-Gaussianity in the weak lensing fields, even if the
primordial fluctuations are Gaussian. This non-Gaussian signal thus
provides additional information on structure formation models that
cannot be extracted by the widely used two-point statistics such as
the power spectrum. The bispectrum, the Fourier counterpart of the
three-point correlation function, is the lowest-order statistical
quantity to describe non-Gaussianity. Future wide-field
surveys promise to measure the bispectrum of lensing fields at high
significance, as will be shown here. The
primary goal  of this paper is to determine the expected accuracy on 
cosmological parameters
from tomography that jointly uses the lensing power
spectrum and bispectrum. The lensing bispectrum or skewness have 
been applied for cosmological parameter estimation 
in previous studies  (Hui 1999; Benabed \& Bernardeau 2001; Cooray \&
Hu 2001a; Refregier et al. 2003). 

In this work, we use the lensing bispectra over all
triangle configurations available from a given survey to compute
the signal-to-noise for the bispectrum measurement. In addition, 
we use all the cross- and auto-bispectra constructed from the lensing
fields in redshift bins. To do this, we correctly take into account the
covariance in the analysis. We study all the parameters that the 
lensing power spectrum and bispectrum are sensitive to and present
results for desired parameters with different marginalization schemes. 

We employ the cold dark matter (CDM) model to
predict the lensing power spectrum and bispectrum for dark energy
cosmologies.  The parameter constraints derived thus depend on the model
of mass clustering (see Jain \& Taylor 2003 and Bernstein \& Jain
2003 for a model-independent method). 
Hence, we derive the
constraints on dark energy parameters marginalized over the other
cosmological parameters on which the lensing observables depend.
On angular scales below a degree, 
non-linear evolution leads to a significant
enhancement of the bispectrum amplitude (compared to perturbation
theory), and is likely to amplify the dark energy
dependences (e.g., Hui 1999; Ma et al. 1999).  
We employ an analytic fitting formula of the mass bispectrum given
by Scoccimarro \& Couchman (2001; see also
Scoccimarro \& Frieman 1999) to make model predictions of the lensing
bispectrum. 
Since the physics involved in weak lensing is only gravity, developing 
an accurate model of the non-linear bispectrum is achievable
from $N$-body simulations, as has been done for
the non-linear power spectrum (e.g., Smith et al. 2003,
hereafter Smith03).

In addition, tomography of the lensing bispectrum offers the
possibility of constraining primordial
non-Gaussianity.  While CMB observations so far have shown that the
primordial fluctuations are close to Gaussian (e.g., Komatsu et
al. 2003), it is still worth exploring primordial non-Gaussianity
from observations of large-scale structure at low redshifts, since
the length scales probed by these two methods are 
different. Exploring primordial non-Gaussianity provides useful
information on  the physics involved in the early
universe, such as particular inflation models (e.g., Wang \& Kamionkowski
1998; Verde et al. 1999; Komatsu 2002; Dvali et al. 2003; 
 Zaldarriaga 2003 and references therein).  The lensing bispectrum due to 
primordial non-Gaussianity has a 
different redshift evolution from that due to the non-Gaussianity of
structure formation and, in addition, their configuration dependences
differ.  Hence, bispectrum tomography can be useful in
separating these two contributions. 
We estimate how measuring the lensing bispectrum from future surveys 
can constrain the primordial non-Gaussianity.

The structure of this paper is as follows. 
In \S \ref{tomo}, we present the formalism for computing the power
spectrum, bispectrum and cross-spectra of the lensing convergence.  The
covariance between the bispectra in redshift bins is derived.  \S \ref{sn}
presents the signal-to-noise ratio for measuring the lensing bispectrum
for future wide-field surveys. From this estimate, we show how 
lensing tomography can put constraints on  primordial non-Gaussian
model. In \S \ref{fisher}, we present the Fisher matrix formalism for
lensing tomography.  In \S \ref{result} 
we present the forecasts for constraints on cosmological
parameters and the equation of state of dark energy.  
We conclude in \S \ref{conc}. 

We will use the  concordance
$\Lambda$CDM model with $\Omega_{\rm cdm}=0.3$, $\Omega_{\rm b}=0.05$,
$\Omega_{\rm de}=0.65$, $n=1$, $h=0.72$ and $\sigma_8=0.9$ as
supported from the WMAP result (Spergel et al. 2003). $\Omega_{\rm
cdm}$, $\Omega_{\rm b}$ and $\Omega_{\rm de}$ are density parameters of
the cold dark matter, baryons and the cosmological constant at
present, $n$ is the spectral index of the primordial power spectrum of
scalar perturbations, $h$ is the Hubble parameter, and $\sigma_8$ is the
rms mass fluctuation in spheres of radius $8h^{-1}$Mpc.

\section{Lensing power spectrum and bispectrum tomography}
\label{tomo}

\subsection{Preliminaries: cosmology}
\label{cosmo}

We work in the context of spatially flat cold dark matter models for
structure formation.  The expansion history of the universe is given by
the scale factor $a(t)$ in a homogeneous and isotropic universe.  The scale
factor during the matter dominated epoch is determined by density
contributions from non-relativistic matter density $\Omega_{\rm m}$ (the
cold dark matter plus the baryons) and dark energy density $\Omega_{\rm
de}$ at present, in units of the critical density $3H_0^2/(8\pi G)$,
where $H_0=100~ h~{\rm km}~ {\rm s}^{-1}~ {\rm Mpc}^{-1}$ is the Hubble
parameter at present. The expansion rate, the Hubble parameter, is given by
\begin{equation}
H^2(a)=H_0^2\left[\Omega_{\rm m}a^{-3}+\Omega_{\rm de}
e^{-3\int^a_1 da' (1+w(a'))/a'},
\right]
\end{equation}
where we have employed the normalization $a(t_0)=1$ today for our
convention and $w(a)$ specifies the equation of state for dark energy as
\begin{equation}
w(a)\equiv \frac{p_{\rm de}}{\rho_{\rm de}}=-\frac{1}{3}
\frac{d\ln \rho_{\rm de}}{d\ln a}-1. 
\end{equation}
Note that $w=-1$ corresponds to a cosmological constant.  The comoving
distance $\chi(a)$ from an observer at $a=1$ to a source at $a$ is
expressed in terms of the Hubble parameter as
\begin{equation}
\chi(a)=\int^1_a\!\!\frac{da'}{H(a')a^{\prime 2}}. 
\end{equation}
This gives the distance-redshift relation $\chi(z)$ via the relation
$1+z=1/a$.

Dark energy that has negative pressure ($\rho_{\rm
de}+3p_{\rm de}<0$ as observed today) leads to repulsive gravity, and
therefore does not cluster significantly. However, dark energy does
affect the growth of mass clustering through its effect on the expansion
rate.  In linear theory, all Fourier modes of the mass density perturbation, 
$\delta(\equiv \delta
\rho_m/\bar{\rho}_m)$, grow at the same rate:
$\tilde{\delta}_{\bm{k}}(a)=D(a)\tilde{\delta}_{\bm{k}}(a=1)$, where
$D(a)$ is the growth factor normalized as $D(a=1)=1$ and, in the
following, the tilde symbol is used to denote the Fourier component. 
The growth factor can be computed by solving the linearized
differential equation,
$\ddot{\delta}_{\bm{k}}+2(\dot{a}/a)\dot{\delta}_{\bm{k}} -4\pi
G\bar{\rho}_m \delta_{\bm{k}}=0$, where the dot is the derivative with
respect to physical time.  Hence, the growth suppression rate (growth
rate relative to that in a flat, matter-dominated universe),
$g(a)=D(a)/a$, can be obtained by solving the differential equation
(e.g., Wang \& Steinhardt 1998):
\begin{eqnarray}
2\frac{d^2g}{d\ln a^2}+\left[5-3w(a)\Omega_{\rm de}(a)\right]
\frac{dg}{d\ln a}+3\left[1-w(a)\right]\Omega_{\rm de}(a)g(a)=0,
\end{eqnarray}
where $\Omega_{\rm de}(a)$ is the dark energy density parameter at epoch
$a$. We employ the initial conditions of $g=1$ and $dg/d\ln a=0$ at
$a_i=1/1100$, which is valid for the dark energy models we consider.
Throughout this paper we assume a spatially smooth dark energy
component (its spatial fluctuations matter on length scales
comparable with the present-day horizon scale, which are not
probed by weak lensing surveys of interest). For example, $g(a)$ for
dark energy cosmologies are shown in Figure 2 of Linder \& Jenkins
(2003). 

In this paper 
we mainly employ the $\sigma_8$ normalization for the input linear mass power
spectrum, equivalent to setting $D(a=1)=1$ today.  
In this case, increasing $w$ from $w=-1$ or equivalently increasing
the dark energy contribution at $z>0$ leads to
slower redshift evolution of the growth rate, since structure
formation freezes at higher redshift; for example
$D_{w=-0.9}(z)>D_{w=-1}(z)$ for $z>0$. 
Hence, the mass power spectrum amplitude is greater for $w>-1$  
than that for the $\Lambda$CDM model at $z>0$. 

\subsection{Weak lensing fields}

\begin{figure}
  \begin{center}
    \leavevmode\epsfxsize=10.cm \epsfbox{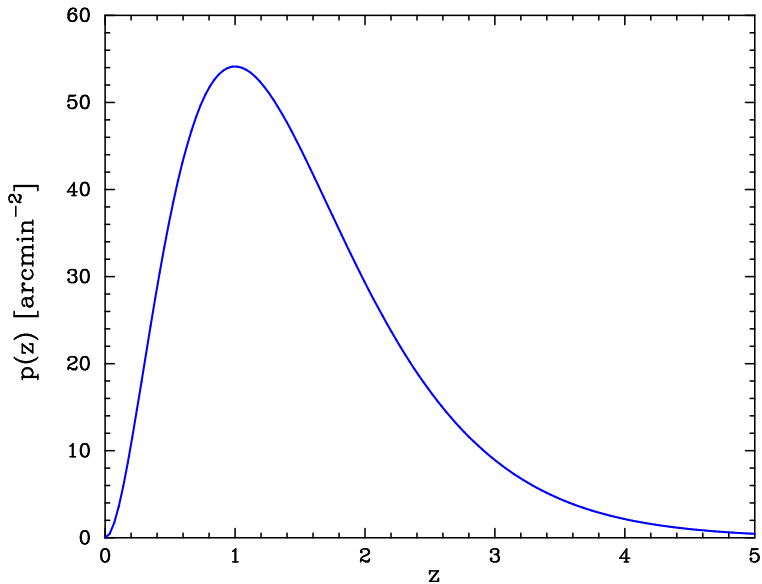}
  \end{center}
\caption{The assumed redshift distribution of source galaxies. The curve
 is in units of the galaxy number density per ${\rm arcmin}^2$ and
 normalized as $\int^\infty_0\!\! dz~ p(z)=\bar{n}_{\rm g}=100 \mbox{
 arcmin}^{-2}$. 
}  \label{fig:pz}
\end{figure}

In the context of cosmological gravitational lensing, the convergence
field is expressed as a weighted projection of the three-dimensional
density fluctuation field between source and observer (e.g., see 
Bartelmann \& Schneider 2001; Mellier 1999 for reviews):
\begin{equation}
\kappa(\bm{\theta})=\int_0^{\chi_H}\!\!d\chi W(\chi) 
\delta[\chi, \chi\bm{\theta}],
\label{eqn:kappa}
\end{equation}
where $\bm{\theta}$ is the angular position on the sky, $\chi$ is the
comoving distance, and $\chi_H$ is the distance to the horizon.  Note
that for a flat universe the comoving angular diameter distance is
equivalent to the comoving distance.  Following 
Blandford et al. (1991), Miralda-Escude (1991) and Kaiser (1992), we have
used the Born approximation, where the convergence field is computed
along the unperturbed path.  The lensing weight function
$W(\chi)$ is defined as
\begin{eqnarray}
W(\chi)=\frac{3}{2}\Omega_{\rm m0}H_0^2 a^{-1}(\chi)
\chi
\frac{1}{\bar{n}_{\rm g}}\int^{\chi_H}_\chi\!\!d\chi_s~ 
p_s(z)\frac{dz}{d\chi_s} 
\frac{\chi_{\rm s}-\chi}{\chi_s},
\label{eqn:weightgl}
\end{eqnarray}
where $p_s(z)$ is the redshift selection function of source galaxies 
and satisfies the normalization condition $\int_0^{\infty}\!\!dz ~ p(z)=
\bar{n}_{\rm g}$, with the average number density per unit steradian
$\bar{n}_{\rm g}$.
Following Huterer (2002), we employ the source distribution given by
\begin{equation}
p_s(z)=\bar{n}_{\rm g}\frac{z^2}{2z_0^3}e^{-z/z_0},
\label{eqn:pz}
\end{equation}
with $z_0=0.5$.  As shown in Figure
\ref{fig:pz}, $p(z)$ peaks at $2z_0=1$ and has
median redshift of galaxies $z_{\rm med}=1.5$. 

Future surveys will provide photometric redshift information on source
galaxies. This additional information allows us to subdivide the galaxies
into redshift bins. The average number density of galaxies in a
redshift bin $i$, defined to lie between the comoving distances 
$\chi_i$ and $\chi_{i+1}$, is given by
\begin{equation}
\bar{n}_i=\int_{\chi_i}^{\chi_{i+1}}\!\!d\chi_s~ p_s(z)\frac{dz}{d\chi_s}. 
\label{eqn:ni}
\end{equation}
Note that this number density determines the shot noise contamination
due to the intrinsic ellipticities of galaxies 
for the power spectrum measurement in the
bin (see equation (\ref{eqn:covps})).  The convergence field for 
subsample $i$ is
\begin{equation}
\kappa_{(i)}\!(\bm{\theta})=\int_0^{\chi_H}\!\!d\chi~ W_{(i)}(\chi) 
\delta[\chi, \chi\bm{\theta}],
\label{eqn:kappai}
\end{equation}
with the lensing weight function $W_{(i)}$ given by
%
\begin{eqnarray}
W_{(i)}(\chi)&=&
\left\{
\begin{array}{ll}
{\displaystyle 
\frac{W_0}{\bar{n}_i}
a^{-1}\! (\chi)~ \chi
\int_{{\rm max}\{\chi,\chi_{i}\}}^{\chi_{i+1}}\!\!d\chi_{s}~ 
p_s(z)\frac{dz}{d\chi_s} \frac{\chi_{\rm s}-\chi}{\chi_s}},
& \chi\le\chi_{i+1},\\
0,& \chi>\chi_{i+1}.
\end{array}
\right.
\label{eqn:weight}
\end{eqnarray}
where $W_0=3/2\ \Omega_{\rm m0}H_0^2$. We have ignored possible errors in
the photometric redshifts for simplicity.
How a dynamically evolving dark energy model changes the lensing weight
function is shown in Figure 3 in Huterer (2002). 
For example, increasing $w$ lowers $W_{(i)}$ --- 
opposite to the dependence of the growth rate of mass clustering 
(for $\sigma_8$ normalization). 
Thus the dependence of lensing observables
on the equation of state is somewhat weakened by these two effects. 
We have checked that the redshift evolution of  the lensing
weight function 
provides the dominant constraints on the equation of state (see also 
Figure 2 in Abazajian \& Dodelson 2003).

\subsection{The power spectrum and its covariance}

To compute the power spectrum and bispectrum of the convergence, we employ
the flat-sky approximation (Blandford et al. 1991; Miralda-Escude 1991;
Kaiser 1992). Within this framework the lensing convergence 
field is decomposed into angular modes based on the two-dimensional Fourier
transform: $\kappa(\bm{\theta})=\sum_{\bm{l}}
\tilde{\kappa}_{\bm{l}}e^{i\bm{l}\cdot\bm{\theta}}$. The angular 
power spectrum, $C(l)$, is defined as
\begin{eqnarray}
\skaco{\tilde{\kappa}_{\bm{l}_1}\tilde{\kappa}_{\bm{l}_2}}
&=&(2\pi)^2\delta^D(\bm{l}_{12})C(l_1),
\end{eqnarray}
where $\delta^D(\bm{l})$ is the Dirac delta function, $\skaco{\cdots}$
denotes ensemble averaging, and
$\bm{l}_{12}=\bm{l}_1+\bm{l}_2$.

\begin{figure}
  \begin{center}
    \leavevmode\epsfxsize=12.cm \epsfbox{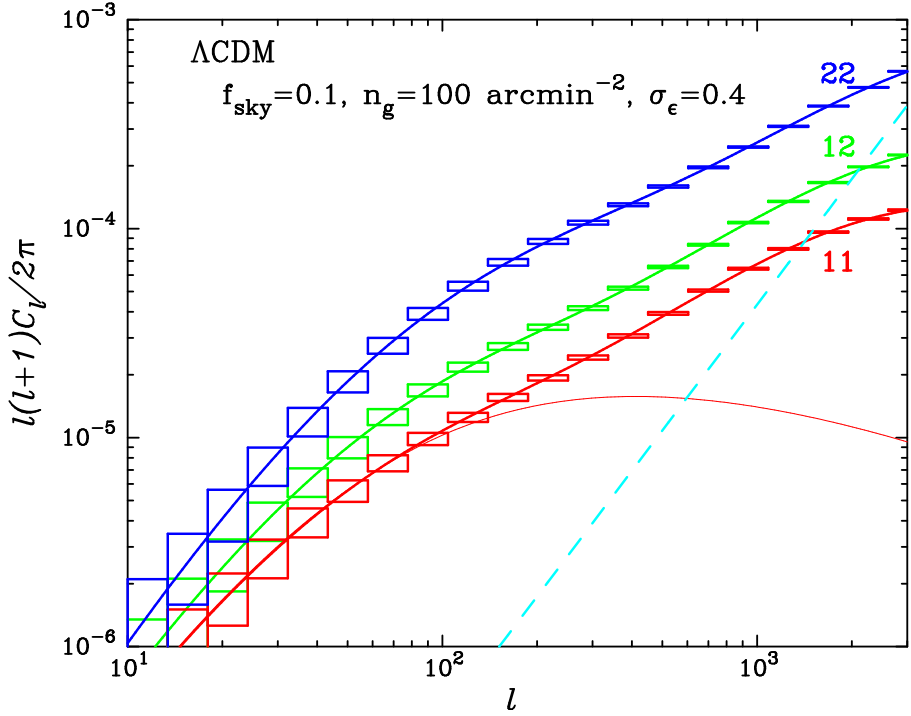}
  \end{center}
\caption{The lensing power spectra constructed from galaxies in two
 redshift bins, $0\le z_1\le 1.3$ and $1.3\le z_2$. 
The solid curves are the results for the fiducial $\Lambda$CDM model, 
computed from the Smith03 fitting formula. The boxes show the expected 
measurement error due to sample variance
and intrinsic ellipticities. The sky coverage is 
$f_{\rm sky}=0.1$ and the rms of intrinsic
 ellipticities, $\sigma_\epsilon=0.4$. 
The linear power spectrum for the $1$ bin 
is shown by the thin solid line to show how significant
the non-linear effect is. The dashed curve shows the shot noise 
contribution to the power spectrum measurement.  }
 \label{fig:cl}
\end{figure}

For lensing tomography, we use all the auto- and cross-power spectra
that are constructed from source galaxies divided into redshift bins.
The angular power spectrum between redshift bins $i$ and $j$, $C_{(ij)}(l)$,
is given by
\begin{eqnarray}
C_{(ij)}(l)
=\int^{\chi_H}_0\!\!d\chi
W_{(i)}\!(\chi) W_{(j)}\!(\chi)
\chi^{-2}~ P_\delta\!\left(k=\frac{l}{\chi}; 
\chi\right),
\label{eqn:cltomo}
\end{eqnarray}
where the lensing weight function $W_{(i)}$ 
is given by equation (\ref{eqn:weight}) 
and $P_\delta(k)$ is the three-dimensional mass power spectrum. 
Using $n_s$ redshift bins
leads to $n_s(n_s+1)/2$ cross and auto power spectra. Note that we
have used Limber's equation (e.g. Kaiser 1992), 
which is a good approximation over the angular modes we consider, 
with $50\le l\le 3000$, corresponding to angular scales between $5'$ and a
few degrees.  
The non-linear gravitational evolution
of $P_\delta(k)$ significantly enhances the amplitude of the lensing
power spectrum on angular scales $l\simgt 100$ 
(see Figure \ref{fig:cl}).  Therefore, we need an accurate model of 
$P_\delta(k)$, for which we employ the fitting formula proposed
by Smith et al. (2003, hereafter Smith03).  We assume that
the Smith03 formula can be applied to dark energy cosmologies, if we
replace the growth factor in the formula with that for a given dark
energy cosmology from which we compute the non-linear scale $k_{\rm NL}$, 
the effective spectral index $n_{\rm eff}$
and the spectral curvature $C$ used in the formula
to obtain the mapping between the linear and non-linear power spectra. 
The issue of accurate power spectra for general dark energy
cosmologies needs to be addressed carefully (see Ma et al. 1999 and 
Huterer 2002 for related discussions).  It is encouraging that
Linder \& Jenkins (2003) find that the mass functions seen in
$N$-body simulations for dark energy cosmologies are well fitted by the
universal fitting formula derived in Jenkins et al. (2001), which is
derived from the same $N$-body simulations as used in Smith03.

Measured shear correlations contain a 
shot-noise contribution from the intrinsic
ellipticities of source galaxies. Assuming that the ellipticity distribution
is uncorrelated between different galaxies, the observed power spectrum
between redshift bins $i$ and $j$ can be expressed as (Kaiser 1992, 1998;
Hu 1999) 
\begin{eqnarray}
C^{\rm obs}_{(ij)}(l)
=C_{(ij)}(l)+\delta_{ij}\frac{\sigma_{\epsilon}^2}{\bar{n}_{i}},
\label{eqn:obscl}
\end{eqnarray}
where $\bar{n}_i$ is the average number density of galaxies in redshift
bin $i$, as given by equation (\ref{eqn:ni}). The Kronecker delta
function enforces the fact that the cross power spectrum with $i\ne
j$ is not affected by shot noise.  In this sense, the
cross power spectrum is an unbiased estimator of the cosmological
signal.  We have ignored other possible contaminations such as 
observational systematics and intrinsic ellipticity alignments (the
latter is also likely to be negligible for cross power spectra as
discussed in Takada \& White (2003)). 

The power spectrum covariance is needed to understand
statistical errors on the power spectrum measurement. 
Extending the expression in Kaiser (1998) and Scoccimarro et al. (1999),
the covariance between the power spectra $C_{(ij)}(l)$ and
$C_{(mn)}(l')$ is 
\begin{eqnarray}
{\rm Cov}[C^{\rm obs}_{(ij)}(l),C^{\rm obs}_{(mn)}(l')]
&=&\frac{\delta_{ll'}}{(2l+1)\Delta l f_{\rm sky}} 
\left[C^{\rm obs}_{(im)}(l)C_{(jn)}^{\rm obs}(l)
+C^{\rm obs}_{(in)}(l)C_{(jm)}^{\rm obs}(l)\right]
\nonumber\\
&& + \frac{1}{4\pi f_{\rm sky}}
\int\!\!\frac{d^2\bm{l}}{A(l)}\int\!\!\frac{d^2\bm{l}'}{A(l')}
T_{(ijmn)}(\bm{l},-\bm{l},\bm{l}',-\bm{l}'), 
\label{eqn:covps}
\end{eqnarray}
where $T_{(ijmn)}$ is the trispectrum of the convergence fields in
redshift bins $i,j,m$ and $n$, $f_{\rm sky}$ is the fraction of sky 
covered and $\Delta l$ is the bin width centered at $l$, so that
the area of the shell is $A(l)=2\pi l \Delta l$. The first term 
denotes the Gaussian contribution to the covariance, and does not 
lead to any correlation between the power spectra of different $l$ 
modes. The second term is
the contribution due to the non-Gaussianity of the weak lensing field.
We restrict our analysis to angular scales $l\le
3000$.  Over this range, the statistical properties of the lensing fields
are quite close to Gaussian (Scoccimarro et al. 1999; White \& Hu 2000; 
Cooray \& Hu 2001b). 
This is physically because the weak lensing field is caused by
independent structures at different redshifts along the line of sight
and this makes the lensing field close to Gaussian due to the central
limit theorem, even though the mass distribution in each redshift slice
displays strong non-Gaussianity.  In what follows, therefore, we
ignore the non-Gaussian contribution to the power spectrum covariance.

Figure \ref{fig:cl} shows the lensing power spectra for two redshift
bins, leading to 3 different power spectra as indicated. The
solid curves are the results from the Smith03 fitting formula. We
parameterized a lensing survey by the sky coverage 
$f_{\rm sky}=0.1$ ($\approx 4000$
degree$^2$), the galaxy number density $\bar{n}_{\rm g}=100~ {\rm
arcmin}^{-2}$ and the rms of intrinsic ellipticities
$\sigma_{\epsilon}=0.4$, respectively. 
 It is clear that the
power spectrum for higher redshift bin has greater amplitude because of
the greater lensing efficiency described by the lensing weight function 
$W_{(i)}$ (see equation (\ref{eqn:weight})).  The correlation coefficient
between the power spectra of the redshift bins,
$R_{ij}=C_{(ij)}(l)/[C_{(ii)}(l)C_{(jj)}(l)]^{1/2}$, quantifies how the
power spectra are correlated. Even with only two
redshift bins, the power spectra are highly correlated as $R_{12}\sim
0.8$. One thus gains little information from further subdivision of the
redshift bins (Hu 1999, 2002a,b).  The box around each curve shows the
expected measurement error at a given bin of $l$, 
which includes the cosmic sample
variance and the error due to 
the intrinsic ellipticities. 
As can
be seen, this type of survey allows the 
power spectrum measurements with significant signal-to-noise ratio.  
The dashed curve shows the shot noise contribution to the power spectrum
measurement. The comparison with the
power spectrum amplitude clarifies that shot noise 
becomes  dominant over or comparable with
sample variance at $l\simgt 10^3$. It is worth noting that
shot noise only contributes to the Gaussian term in the power
spectrum covariance, thus
strengthening the case for our Gaussian error
assumption (Cooray \& Hu 2001b).
Finally, to clarify the effect of the non-linear gravitational clustering, 
the thin solid curve shows the prediction of $C_{(11)}(l)$ from the
linear mass power spectrum: the non-linear effect induces a significant
enhancement in the amplitude by about an order of magnitude at
$l\simgt 100$.

\subsection{Convergence bispectrum and its covariance}

\subsubsection{Definition}

In this subsection, we consider tomography with the lensing
bispectrum, which is the main focus of this paper.  The bispectrum is the
lowest-order quantity to extract non-Gaussianity in the weak lensing
field, and thus provides additional information on structure formation
models relative to the power spectrum (e.g., Takada \& Jain 2002,
2003a,b,c, hereafter TJ02 and TJ03a,b,c).  For convenience of the
following discussion,
we start with the definition
of the bispectrum based on the harmonic expansion of the
convergence field that accounts for the curvature of a celestial sphere
(Hu 2000; Komatsu 2002):
$\kappa(\bm{\theta})=\sum_{lm}\kappa_{lm}Y_{lm}(\bm{\theta})$.  The
bispectrum, $B_{(ijk)l_1l_2l_3}$, between the convergence fields in
redshift bins $i$, $j$ and $k$, is defined as
\begin{eqnarray}
\skaco{\kappa_{l_1m_1(i)}\kappa_{l_2m_2(j)}
\kappa_{l_3m_3(k)}}
=\left(
\begin{array}{lll}
l_1&l_2&l_3\\
m_1&m_2&m_3
\end{array}
\right)B_{(ijk)l_1l_2l_3},
\label{eqn:fullbisp}
\end{eqnarray}
where $\left(
\begin{array}{ccc}
l_1&l_2&l_3\\
m_1&m_2&m_3
\end{array}
\right)$ is the Wigner-3$j$ symbol. The triangle condition of
$|l_i-l_j|\le l_k\le l_i+l_j$ is imposed for all
($m_1,m_2,m_3$), and $l_1+l_2+l_3={\rm even}$ is required from
statistical parity invariance of the angular correlation.

The relation between the bispectra of the all-sky approach and the
flat-sky approximation 
is given by
\begin{eqnarray}
B_{(ijk)l_1l_2l_3}\approx
\left(
\begin{array}{lll}
l_1&l_2&l_3\\
0&0&0
\end{array}
\right)
\sqrt{\frac{(2l_1+1)(2l_2+1)(2l_3+1)}{4\pi}}
 B_{(ijk)}(\bm{l}_1,\bm{l}_2,\bm{l}_3).
\label{eqn:relbisp}
\end{eqnarray}
We will use this equation to compute the all-sky lensing
bispectrum from the flat-sky bispectrum (see below), with
the Wigner-3$j$ symbol, given by the approximation described in Appendix A.

For a given cosmological model, we use the flat-sky approximation to predict
the bispectrum since it
is sufficiently accurate over
angular scales of our interest. Combining the flat-sky
approximation and Limber's equation leads to a simple expression for
the bispectrum in redshift bins $i$,$j$ and $k$,
$B_{(ijk)}(\bm{l}_1,\bm{l}_2,\bm{l}_3)$:
\begin{equation}
\skaco{\tilde{\kappa}_{(i)}(\bm{l}_1)\tilde{\kappa}_{(j)}(\bm{l}_2)
\tilde{\kappa}_{(k)}(\bm{l}_3)}=(2\pi)^2
B_{(ijk)}(\bm{l}_1,\bm{l}_2,\bm{l}_3)\delta^D(\bm{l}_{123})
\label{eqn:fbisp}
\end{equation}
with
\begin{eqnarray}
B_{(ijk)}(\bm{l}_1,\bm{l}_2,\bm{l}_3)=\int^{\chi_H}_0\!\!d\chi
W_{(i)}\!(\chi)W_{(j)}\!(\chi)W_{(k)}\!(\chi)
\chi^{-4}~ B_\delta\!\left(\bm{k}_1,\bm{k}_2,\bm{k}_3;
\chi\right), 
\label{eqn:fbispdef}
\end{eqnarray}
where $\bm{k}_i=\bm{l}_i/\chi$ and
$B_\delta(\bm{k}_1,\bm{k}_2,\bm{k}_3)$ is the bispectrum of the
three-dimensional mass fluctuations.  Note that
the delta function 
$\delta^D(\bm{l}_{123})$
in the bispectrum (\ref{eqn:fbisp}) enforces the condition 
that the three vectors $\bm{l}_1$, $\bm{l}_2$ and
$\bm{l}_3$ form a triangle configuration in the Fourier space.
From statistical isotropy, the convergence
bispectrum is a function of the three parameters that specify the
triangle configuration, e.g., $l_1, l_2$ and $l_3$. 
For $n_s$ redshift bins, $n_s^3$ lensing bispectra contribute to the Fisher
matrix for a given set of (unequal) $(l_1,l_2,l_3)$
\footnote{The number of bispectra is reduced to $n_s(n_s+1)/2$
and $n_s(n_s+1)(n_s+2)/6$ for isosceles triangles, e.g. $l_1=l_2\ne
l_3$, and equilateral triangles with $l_1=l_2=l_3$, respectively.}. 
For comparison, there are $n_s(n_s+1)/2$ power spectra. 
Note that equations (\ref{eqn:fbisp}) and
(\ref{eqn:fbispdef}) show there are symmetry relations among the $n_s^3$
bispectra in an ensemble average sense: e.g., for $n_s=2$ we have the
relations  $B_{(112)}=B_{(121)}=B_{(211)}$ and
$B_{(122)}=B_{(212)}=B_{(221)}$,
but they are different in the sense that their estimators and covariances
are different, as shown below.

In the following, we consider two sources for the mass bispectrum
$B_\delta$: one is primordial
non-Gaussianity, which could be imprinted in the early universe,
and the other is the
non-Gaussianity induced by non-linear gravitational clustering 
from primordial (nearly) Gaussian fluctuations.

\subsubsection{Bispectrum from primordial non-Gaussianity}
\label{pNG}

Following Verde et al. (1999), we consider a model in which the
primordial gravitational potential $\Phi(\bm{x})$ 
is a linear combination of a Gaussian random field
$\phi$ and a term proportional to the square of the same random field:
\begin{equation}
\Phi(\bm{x})=\phi(\bm{x})
+f_{\rm NL}\left[\phi^2(\bm{x})-\skaco{\phi^2(\bm{x})}\right]
\label{eqn:pNG}
\end{equation}
where $f_{\rm NL}$ parameterizes the non-Gaussian amplitude: in the
limit $f_{\rm NL}\rightarrow 0$, the field becomes Gaussian.  The
physical motivation for this model (given that there is a range of possible
non-Gaussian models) is that such a non-Gaussian field naturally arises in
slow-roll inflation and other inflation models (Luo 1994; Falk et
al. 1993; Gangui et al. 1994, and also see Komatsu 2001 for a review) or
in models in which the epoch of reheating has a spatial dependence 
(Dvali et al. 2003; Zaldarriaga 2003).  The
Sachs-Wolfe effect measured in the CMB power spectrum implies $\Phi\sim
10^{-5}$ and the potential remains constant in time on horizon scales. 
From this amplitude and the current limit on $f_{\rm NL}<100$
from the CMB bispectrum measurement (Komatsu et al. 2003),
the second term in equation (\ref{eqn:pNG}) is
much smaller than the first term. As long as $\Phi\ll 1$ at a given
length scale, the redshift evolution of $\Phi$ obeys 
linear theory. 

A future wide-field lensing survey will enable accurate measurements
of  the lensing bispectrum down to arcminute
scales,  as shown below. On scales larger than a degree, the lensing
convergence arises from structures in the linear or weakly nonlinear
regime, so that different Fourier modes are almost uncorrelated. 
Hence the lensing field retains the imprint of primordial
non-Gaussianity. On smaller scales, it remains to be seen how 
highly non-linear structures show signatures from
primordial non-Gaussianity (see Matarrese et al. 2000 for an 
analytic discussion). 

To consider the effect of primordial non-Gaussianity on  weak
lensing fields, we need to account for the different evolution of modes
inside the horizon.  We assume that, in the linear regime, the evolution can
be described by the transfer function. Since the density fluctuation
field is related to the potential field via the Poissonian equation, the
Fourier transform of the density fluctuation field, 
$\tilde{\delta}(\bm{k})$, at epoch $z$ can be expressed as
\begin{equation}
\tilde{\delta}(\bm{k};z)=D(z) {\cal M}_k\tilde{\Phi}(\bm{k}),
\end{equation}
with 
\begin{equation}
{\cal M}_k=-\frac{2}{3H_0^2\Omega_{\rm m0}}k^2T(k),
\label{eqn:kernelMk}
\end{equation}
where $D(z)$ is the linear growth rate and 
$T(k)$ is the transfer function as given e.g. by 
the BBKS formula (Bardeen et al. 1986). We assume that the
power spectrum of $\tilde{\Phi}$ is given by a single power law,
$P_\Phi(k)\propto k^{n-4}$, and the amplitude is
typically $k^3P_\Phi\sim 10^{-5}$ as stated above.
Hence, the linear mass power spectrum at redshift $z$ is given by
$P^L_\delta(k;z)=D(z)^2{\cal M}_k^2P_\Phi(k)$. 
For a plausible range of $f_{\rm NL}$, the
non-Gaussian contribution to the mass power spectrum is much
smaller than the Gaussian contribution as stated above, and
therefore it is a good approximation to fix the normalization of
$P^L_\delta(k)$ from the Gaussian term alone, based on 
measurements such as the CMB anisotropy or the $\sigma_8$
normalization.

Even in the linear regime, the bispectrum of mass fluctuations 
due to primordial non-Gaussianity  at redshift $z$ is given by
\begin{eqnarray}
B_{\delta}^{\rm NG}(\bm{k}_1,\bm{k}_2,\bm{k}_3; z)
=2f_{\rm NL}D^3(z) \left[
P^L_{\delta}(k_1)P^L_{\delta}(k_2)\frac{{\cal M}_{k_3}}
{{\cal M}_{k_1}{\cal M}_{k_2}}
+2 \mbox{perm.}\right]. 
\label{eqn:pbisp}
\end{eqnarray}
Substituting this expression into equation (\ref{eqn:fbispdef}) gives
the prediction for the lensing bispectrum. We will use the prediction to 
estimate how a wide-field lensing survey can constrain $f_{\rm NL}$.

\subsubsection{Bispectrum due to gravitational clustering}
\label{LSS}

Even if the primordial fluctuations are Gaussian, non-linear
gravitational clustering  at low redshifts leads
to non-Gaussian features in the mass distribution due to the 
coupling of different Fourier modes. 

In the weakly non-linear regime ($\delta\simlt 1$), the evolution of
mass clustering can be described by perturbation theory (e.g.,
Bernardeau et al. 2002). Perturbation theory ceases to be accurate
in the non-linear regime $(\delta\simgt 1)$. Non-linear clustering 
enhances the amplitude of mass fluctuations which significantly
increases the lensing signal on angular scales below a degree, 
as shown in Figure \ref{fig:cl} (see also Jain \& Seljak 1997).
Therefore, it is necessary to account for non-linear effects to 
correctly interpret measurements from lensing surveys. 
However, an accurate model of the non-linear mass bispectrum is not
yet available, in part because simulations needed to calibrate 
such model predictions are challenging to carry out: many realizations of
high resolution simulations are needed to calibrate the 
triangle configuration dependence of the bispectrum. 
For this study we will use the best available models to 
clarify the usefulness of bispectrum tomography. While the resulting 
parameter accuracy forecasts are likely to be adequate, to actually 
infer the correct values of parameters from measurements, better
predictions will be needed. 

There are two well studied analytic models of nonlinear clustering 
that we have implemented: hyper-extended 
perturbation theory (Scoccimarro \& Frieman 1999; Scoccimarro \&
Couchman 2001) and the dark
matter halo approach (e.g., Cooray \& Hu 2001a, TJ03b,c, Takada \&
Hamana 2003  and Cooray \& Sheth 2002 for a review). 
These models are accurate at the $10-30\%$ level 
in the amplitude of the three-point correlation function (TJ03b,c).
In this paper, we employ the fitting formula of Scoccimarro \&
Couchman for the mass bispectrum, since it makes it feasible to
evaluate the Fisher matrix of bispectrum tomography 
with reasonable computational expense compared to the halo
model implementation: 
\begin{eqnarray}
B^{\rm GRAV}_\delta(\bm{k}_1,\bm{k}_2,\bm{k}_3; z)
&=&2 F_2^{\rm eff}(\bm{k}_1,\bm{k}_2)
P^{NL}_\delta(k_1;z)P^{NL}_\delta(k_2;z)+2{\rm ~perm.},
\label{eqn:bisp}
\end{eqnarray}
with the kernel $F_2^{\rm eff}$ given by
\begin{equation}
F_2^{\rm eff}(\bm{k}_1,\bm{k}_2)=\frac{5}{7}a(n_{\rm eff},k_1)a(n_{\rm eff},k_2)
+\frac{1}{2}\left(\frac{k_1}{k_2}+\frac{k_2}{k_1}\right)
\frac{(\bm{k}_1\cdot\bm{k}_2)}{k_1k_2}b(n_{\rm eff},k_1)b(n_{\rm eff},k_2)
+\frac{2}{7}\frac{(\bm{k}_1\cdot\bm{k}_2)^2}{k_1^2k_2^2}c(n_{\rm eff},k_1)c(n_{\rm eff},k_2). 
\end{equation}
Note that $P^{NL}_\delta$ in equation (\ref{eqn:bisp}) is the non-linear
mass power spectrum for which we use the Smith03 fitting formula. The
fitting functions in $F_2^{\rm eff}$ are given by 
\begin{eqnarray}
a(n_{\rm eff},k)&=&\frac{1+\sigma_8^{-0.2}(z)\sqrt{0.7Q_3(n_{\rm eff})}(q/4)^{n_{\rm eff}+3.5}}
{1+(q/4)^{n_{\rm eff}+3.5}},\nonumber\\
b(n_{\rm eff},k)&=&\frac{1+0.4(n_{\rm eff}+3)q^{n_{\rm eff}+3}}
{1+q^{n_{\rm eff}+3.5}},\nonumber\\
c(n_{\rm eff},k)&=&\frac{1+\left(\frac{4.5}{1.5+(n_{\rm eff}+3)^4}\right)
(2q)^{n_{\rm eff}+3}}
{1+(2q)^{n_{\rm eff}+3.5}},
\end{eqnarray}
where $n_{\rm eff}$ is the effective spectral index of the power
spectrum at the scale $k$ defined as $n_{\rm eff}(k)\equiv
d\ln P_\delta^{L}/d\ln k$ ($P^L_\delta(k)$ is the linear mass power spectrum). 
The quantities $q$ and $Q_3$ are defined by $q=k/k_{\rm NL}$ with
$(k^3/2\pi^3)D^2(z)P^L_\delta(k_{\rm NL})=1$ and $Q_3(n_{\rm
eff})=(4-2^{n_{\rm eff}})/(1+2^{n_{\rm eff}+1})$, respectively, and
$\sigma_8(z)=D(z)\sigma_8$. 

\begin{figure*}
  \begin{center}
    \leavevmode\epsfxsize=17.cm \epsfbox{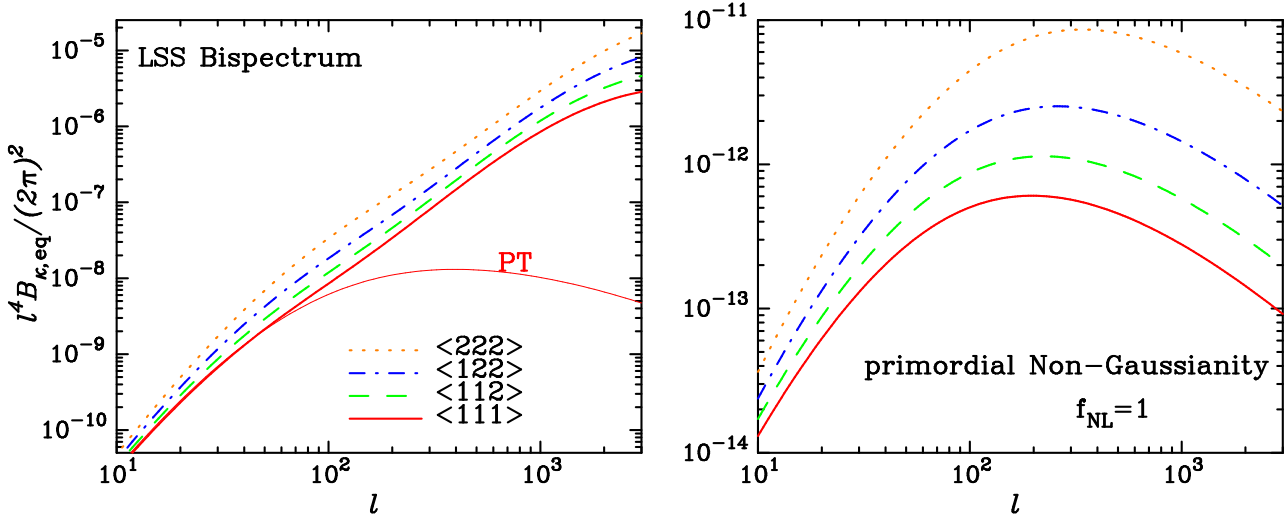}
  \end{center}
\caption{{\em Left panel}: The convergence bispectra
with two redshift bins for equilateral triangles, plotted 
against triangle side length. 
For two redshift bins, there are 4 different bispectra, as
indicated. For comparison
the thin solid curve shows the perturbation
 theory prediction for $B_{(111)}$. Non-linear clustering 
significantly enhances the bispectrum amplitude for $l\simgt 100$. 
{\em Right panel}: Similar plot, but for the bispectra due to 
primordial non-Gaussianity, which are computed from equation
(\ref{eqn:pbisp}). The non-Gaussian parameter $f_{\rm NL}$ is taken to
be $1$. 
} \label{fig:bispns2}
\end{figure*}

One can find that for large length scales, where $k\ll k_{\rm NL}$
(or equivalently $q\ll 1$), $a=b=c=1$ and the bispectrum
(\ref{eqn:bisp}) recovers the tree level perturbation theory
prediction. On the other hand,  on small scales, $k\gg k_{\rm NL}$,
$a=\sigma_8^{-0.2}(z)\sqrt{0.7Q_3(n_{\rm eff})}$ and $b=c=0$, so the
bispectrum becomes independent of triangle configuration and thus
provides the
so-called hierarchical ansatz.  However, it should be noted that a more
realistic non-linear clustering likely displays weak violation of
the hierarchical ansatz in the strongly non-linear regimes (TJ03b,c). 

It is worth mentioning about how the bispectrum (\ref{eqn:bisp}) depends
on equation of state of dark energy. 
In the weakly non-linear regime, the dark energy dependence enters into
the growth factor only and it does not modify the configuration dependence
(as long as we ignore the spatial fluctuations that
modify the transfer function on horizon scales, as shown in Caldwell et
al. 1998, Ma et al. 1999). In the non-linear regime, the
dependence is captured in a complex way through the non-linear power spectrum
 and $\sigma_8(z)$ and the non-linear scale $k_{\rm
NL}(z)$ in the fitting functions $a$, $b$ and $c$. However,
the dark energy
dependence on the {\em lensing bispectrum}, obtained after projecting 
the mass bispectrum with lensing efficiency, 
is mainly determined by the
dependence of the lensing weight function (\ref{eqn:weightgl}).
 
A comparison of equations (\ref{eqn:pbisp}) and (\ref{eqn:bisp})
shows the differences between the bispectra due to the primordial
non-Gaussianity and gravitational clustering. 
The first difference 
is in the redshift dependence: the bispectrum from 
primordial non-Gaussianity has a slower
redshift evolution than the gravitational clustering bispectrum. Thus
lensing tomography should be effective in separating these
two contributions.  The bispectra also differ in the dependence on triangle
configuration: the gravity bispectrum has a stronger
configuration dependences on large
length scales where the primordial non-Gaussianity is relevant.

Figure \ref{fig:bispns2} shows the convergence bispectra 
with two redshift bins against side length $l $ of 
equilateral triangles, as in Figure \ref{fig:cl}.  
The left panel shows the model predictions from the hyper extended
perturbation theory (\ref{eqn:bisp}),
while the right panel shows the
predictions from the primordial non-Gaussian model (\ref{eqn:pbisp})
with $f_{\rm NL}=1$. For two redshift bins, we have 4 different lensing
bispectra as indicated.  For comparison, the thin curve in the left
panel shows the lensing bispectrum computed from the perturbation theory
of structure formation. It is obvious that the non-linear
gravitational clustering significantly enhances the bispectrum amplitude
relative to the perturbation theory prediction by more than an order of
magnitude at $l\simgt 100$. Note
that the halo model used in TJ03c
provides the similar bispectra to within $25\%$ difference in the
amplitude over the range of angular scales we consider.  
From the right panel,  one can find that 
the amplitude of the bispectrum $B_\kappa^{\rm
NG}$ from the primordial non-Gaussian model is much smaller than that 
from the structure formation bispectrum $B_\kappa^{\rm GRAV}$ over the
range of angular scales we consider (this is
true even if we use the current upper limit $f_{\rm NL}\simeq 100$ in
Komatsu et al. 2003).  

\subsubsection{Bispectrum covariance}

In this subsection, we derive the covariance of the convergence
bispectrum.
To do this, we start with considering an estimator of the bispectrum. 
 One practical advantage for measuring the bispectrum (and more
generally odd numbered correlation functions) is that the intrinsic
ellipticities do not contaminate the measurement, as long as the
intrinsic ellipticity distribution is symmetric with zero mean. 
Thus the measured bispectrum is an unbiased estimator of the
cosmological signal:
\begin{equation}
B^{\rm obs}_{(ijk)}(\bm{l}_1,\bm{l}_2,\bm{l}_3)\approx B_{(ijk)}
(\bm{l}_1,\bm{l}_2,\bm{l}_3). 
\end{equation}
It is also worth noting that the $B$-mode induced by intrinsic
ellipticity alignment does not contaminate the bispectrum measurement 
due to statistical parity invariance (Schneider 2003).

\begin{figure}
  \begin{center}
    \leavevmode\epsfxsize=9.cm \epsfbox{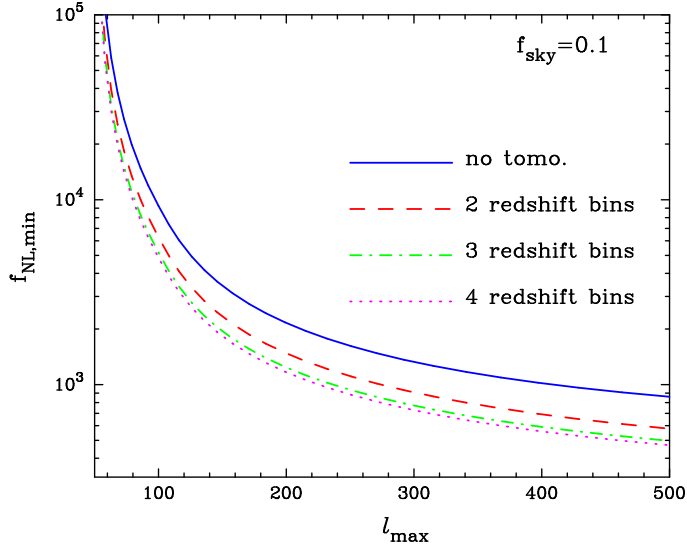}
  \end{center}
\caption{The minimum value of $f_{\rm NL}$ in the primordial
 non-Gaussian model 
such that the lensing bispectrum is detectable with $S/N\ge
 1$, for a survey with $f_{\rm sky}=0.1$. It is plotted 
against the maximum multipole moment $l_{\rm max}$. Here we 
ignore the contribution from nonlinear gravitational
clustering. Note that $f_{\rm NL}$ scales as
 $f_{\rm sky}^{-1/2}$. The different curves show the results for
varying numbers of redshift bins used. 
}  \label{fig:fnl}
\end{figure}

Extending the expression of the bispectrum covariance derived within the
full-sky approach (Hu 2000), the covariance between the lensing
bispectra (\ref{eqn:fullbisp}) in redshift bins can be simply expressed
as
\begin{eqnarray}
{\rm Cov}\left[B_{l_1l_2l_3(abc)},
B_{l_1^\prime l_2^\prime l_3^\prime(ijk)}\right]
&\approx& 
C_{(ai)}^{\rm obs}(l_1)\delta_{l_1l_1^{\prime}}
\left[
C_{(bj)}^{\rm obs}(l_2)
C_{(ck)}^{\rm obs}(l_3)\delta_{l_2l_2^{\prime}}\delta_{l_3l_3^{\prime}}
+C_{(bk)}^{\rm obs}(l_2)
C_{(cj)}^{\rm obs}(l_3)\delta_{l_2l_3^{\prime}}\delta_{l_3l_2^{\prime}}
\right]\nonumber\\
&+&
C_{(aj)}^{\rm obs}(l_1)\delta_{l_1l_2^{\prime}}
\left[
C_{(bi)}^{\rm obs}(l_2)
C_{(ck)}^{\rm obs}(l_3)\delta_{l_2l_1^{\prime}}\delta_{l_3l_3^{\prime}}
+C_{(bk)}^{\rm obs}(l_2)
C_{(ci)}^{\rm obs}(l_3)\delta_{l_2l_3^{\prime}}\delta_{l_3l_1^{\prime}}
\right]\nonumber\\
&+&
C_{(ak)}^{\rm obs}(l_1)\delta_{l_1l_3^{\prime}}
\left[
C_{(bi)}^{\rm obs}(l_2)
C_{(cj)}^{\rm obs}(l_3)\delta_{l_2l_1^{\prime}}\delta_{l_3l_2^{\prime}}
+C_{(bj)}^{\rm obs}(l_2)
C_{(ci)}^{\rm obs}(l_3)\delta_{l_2l_2^{\prime}}\delta_{l_3l_1^{\prime}}
\right].
\label{eqn:covbisp}
\end{eqnarray}
For finite sky coverage, the bispectrum covariance scales as $f_{\rm
sky}^{-1}$.  We have ignored the non-Gaussian contributions to the
covariance that arise from the connected 3-, 4- and 6-point functions of
the convergence field -- see discussion for the power spectrum
following equation (\ref{eqn:covps}).  This is in part verified by examining
the ratio of the non-Gaussian
to the Gaussian contribution to the covariance. 
For example, this ratio for one of these non-Gaussian terms is given 
by $B^2/C_l^3$, and Figures \ref{fig:cl} and \ref{fig:bispns2} show
 that this is much smaller than unity over angular scales of interest.
The Kronecker
delta functions such as $\delta_{l_1l_1^{\prime}}
\delta_{l_1l_2^{\prime}}\delta_{l_1l_3^{\prime}}$ in the equation above
guarantee that the bispectra of different triangles are uncorrelated,
which makes the Fisher matrix analysis significantly simplified. Note
that, on the
other hand,  the bispectra in different redshift bins are highly correlated.

In the following Fisher matrix analysis, we impose the condition $l_1\le
l_2\le l_3$ so that every triangle configuration is counted once. This
condition leads to the following expression for the covariance:
\begin{equation}
{\rm Cov}\left[B_{l_1l_2l_3(abc)},
B_{l_1^\prime l_2^\prime l_3^\prime(ijk)}\right]
\approx  \Delta(l_1,l_2,l_3) f_{\rm sky}^{-1}
C_{(ai)}^{\rm obs}(l_1)C_{(bj)}^{\rm obs}(l_2)
C_{(ck)}^{\rm obs}(l_3)\delta_{l_1l_1^\prime}
\delta_{l_2l_2^\prime}\delta_{l_3l_3^\prime},
\label{eqn:covbisp2}
\end{equation}
where $\Delta(l_1,l_2,l_3)=1$ if all $l$'s are different,
$\Delta(l_1,l_2,l_3)=2$ if two $l$'s are repeated and
$\Delta(l_1,l_2,l_3)=6$ if $l_1=l_2=l_3$, respectively. The observed
power spectrum is given by equation (\ref{eqn:obscl}).  Strictly
speaking, although
equation (\ref{eqn:covbisp2}) is correct only for  $l_1\ne
l_2\ne l_3$
when we consider redshift bins, we checked that
it is a good approximation for the Fisher matrix analysis that
follows, because triangles with $l_1\ne l_2\ne l_3$ 
provide the dominant contribution over angular scales
of interest. 
 Finally, equation (\ref{eqn:covbisp2}) shows that 
for two redshift bins we have the equality
$B_{(112)}=B_{(121)}=B_{(211)}$ (similar conditions hold
for other indices) as stated below equation
(\ref{eqn:fbispdef}), but their covariances are indeed 
different for a general set of ($l_1,l_2,l_3$). 

\begin{figure}
  \begin{center}
    \leavevmode\epsfxsize=9.cm \epsfbox{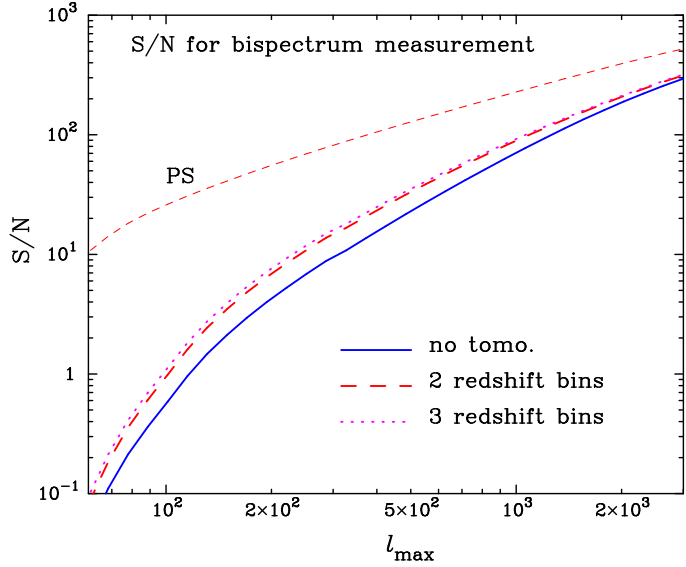}
  \end{center}
\caption{Signal-to-noise ratio estimate for measuring the lensing
 bispectrum against the maximum multipole moment $l_{\rm max}$, as in
 the previous figure. We consider the bispectrum due to structure
 formation and ignore the contribution from primordial
 non-Gaussianity.  For comparison, the thin dashed curve shows the
 signal-to-noise for the power spectrum for 2 redshift bins. 
It is clear
 that $l_{\rm max}^3$ triangles allow significant $S/N$ for the bispectrum 
measurement comparable with that for power
 spectrum at $l_{\rm max}\simgt 1000$. 
 } \label{fig:snbisp}
\end{figure}

\section{Signal-to-noise for the bispectrum} \label{sn}

Before going to parameter accuracy forecasts, we examine how feasible it
is to simply measure a non-zero convergence bispectrum from a wide-field
lensing survey.  Given the bispectrum covariance (\ref{eqn:covbisp2}), the
cumulative signal-to-noise ($S/N$) ratio for measuring the bispectra
in redshift bins can be expressed as
\begin{equation}
\left(\frac{S}{N}\right)^2=\sum_{l_{\rm min}\le 
l_1\le l_2\le l_3\le l_{\rm max}}
\sum_{(i,j,k),(a,b,c)}B_{l_1l_2l_3(ijk)}
\left[{\rm Cov}[B_{l_1l_2l_3(ijk)},B_{l_1l_2l_3(abc)}]
\right]^{-1}B_{l_1l_2l_3(abc)},
\label{eqn:snbisp}
\end{equation}
where $[{\rm Cov}]^{-1}$ denotes the inverse of the covariance matrix,
and $l_{\rm min}$ and $l_{\rm max}$ denote the minimum and maximum
multipole moments considered. The condition $l_1\le l_2\le
l_3$ is imposed, as stated above.
Each of the labels ($a,b,c$) and $(i,j,k)$ denote redshift bins running
over $1,\cdots,n_s$, thus providing contributions from $n_s^3$ bispectra
to the $S/N$.  Given a survey that probes $l$ up to 
$l_{\rm max}$, a rough estimate of the $S/N$ 
is
\begin{eqnarray}
\left(\frac{S}{N}\right)^2&\sim& 
f_{\rm sky}l_{\rm max}^6
\left(
\begin{array}{lll}
l_{\rm max}&l_{\rm max}&l_{\rm max}\\
0&0&0
\end{array}
\right)^2 
\frac{B(l_{\rm max})^2}{C(l_{\rm max})^3}
\sim f_{\rm sky}l_{\rm max}^2 Q^2_\kappa [l_{\rm max}^2C(l_{\rm max})]
\nonumber\\
&\sim& 10^5 \left(\frac{f_{\rm sky}}{0.1}\right)
\left(\frac{l_{\rm max}}{10^3}\right)^2
\left(\frac{Q_\kappa}{100}\right)^2
\left(\frac{l_{\rm max}^2C(l_{\rm max})}{10^{-4}}\right),
\label{eqn:snapp}
\end{eqnarray}
where we have used equation (\ref{eqn:relbisp}) and 
$\sum_{l_1,l_2,l_3\le l_{\rm max}}\sim l_{\rm
max}^3$, $\left(\begin{array}{lll} l&l&l\\ 0&0&0
\end{array}
\right)\sim l^{-1}$ for $l\gg 1$ and introduced the reduced bispectrum
$Q_\kappa$ defined as $Q_\kappa \sim B(l)/[C(l)]^2$, often used in the
literature (e.g., TJ03c).  
Note that in the estimate above we 
ignored the shot noise contribution of
intrinsic ellipticities to the covariance, and thus the estimate
overestimates the $S/N$ (see Figure \ref{fig:snbisp}).
As implied from Figures \ref{fig:cl} and \ref{fig:bispns2} (also see 
Figure 6 in TJ03c), a plausible model for the convergence
bispectrum leads to $60\simlt
Q_\kappa\simlt 150$ for $l\simgt 100$ (corresponding to $\theta\simlt 1$
degree) for the $\Lambda$CDM model. This rough estimate shows that a
future survey with $f_{\rm sky}=0.1$ would allow for a detection
with $S/N\sim 300$ for $l_{\rm max}=10^3$, if we combine the information
from all triangle configurations available.

We give below a more quantitative $S/N$ estimate for measuring the
convergence bispectrum.  As described in \S \ref{pNG} and \ref{LSS},
there are two cosmological contributions to the lensing bispectrum,
due to the primordial non-Gaussianity and gravitational clustering. 
The observed bispectrum can be
expressed as $B_\kappa=B_{\kappa}^{\rm NG}+B_\kappa^{\rm GRAV}$.

In Figure \ref{fig:fnl} we examine the smallest $f_{\rm NL}$
against the maximum multipole moment $l_{\rm max}$, defined so that
the lensing bispectrum is
detectable with $S/N\ge 1$. 
We compute this $f_{\rm NL}$ by setting $B_\kappa=B^{\rm NG}_\kappa$
(that is, we ignore the contamination from the bispectrum induced by
gravitational clustering $B_\kappa^{\rm GRAV}$).  The sky coverage is
taken to be $f_{\rm NL}=0.1$ and the minimum multipole moment 
$l_{\rm min}=50$ so that 
$l_{\rm min}\sim 10 \times 2\pi/\sqrt{4\pi f_{\rm sky}}$.
The figure shows that the tomography of the lensing bispectrum can lower
the $f_{\rm NL}$ value, reflecting the fact that $n_s^3$ 
bispectra of the tomography  provide additional information
on the $S/N$.  As can be seen, $f_{\rm NL}$ is mainly determined by 
modes with $l\simlt 100$, due to the high-$k$ suppression of the kernel
${\cal M}_{k}$ in the 3D mass bispectrum 
(see equation (\ref{eqn:kernelMk}) and Figure 1 in Verde et al. 1999).
In addition, at higher $l$ 
the structure formation bispectrum is more dominant 
(see Figure \ref{fig:bispns2}).  Therefore, focusing 
on large angular scales $\simgt 1$ degree is probably an  appropriate 
strategy for constraining primordial non-Gaussianity. 

The strongest constraint is $f_{\rm NL}\approx 150 f_{\rm sky}^{-1/2}$ for
$l_{\rm max}=500$ and four redshift bins. This result is comparable to
the forecast from the galaxy bispectrum measurement from the SDSS/2dF
surveys in Verde et al. (1999), where the range $f_{\rm NL}=10^3-10^4$
was derived (for $f_{\rm sky}=10^{-5}-0.25$ for galaxy catalogs).
Unless an almost all-sky lensing survey is available,
the CMB bisepctrum is likely to provide more stringent constraint on
$f_{\rm NL}$, as shown by the WMAP result, $-59<f_{\rm
NL}<134$ at $95\%$ confidence, in Komatsu et al. (2003). Hence, we
will ignore the primordial non-Gaussian contribution in the following
analysis. Only if a significantly deeper all sky lensing survey 
is feasible, would it be worth returning to this question and exploring
measurement strategy. 
 
Figure \ref{fig:snbisp} shows the signal-to-noise for measuring
the lensing bispectrum due to structure formation, $B_\kappa^{\rm
GRAV}$.  Note that for $l_{\rm max}\ge 300$ 
we used a binning of $l$ in
the $S/N$ evaluation as described below (see equation
(\ref{eqn:fisherbin})), since the direct summation over $l$'s is
computationally time-consuming for large $l_{\rm max}$.  
It is obvious that a survey with $f_{\rm sky}=0.1$
allows a significant detection of the lensing bispectrum with
$S/N\simgt 10$ for $l_{\rm max}\simgt 300$.  
Bispectrum tomography leads to only slight
improvement in the $S/N$ value.
For comparison, the top thin curve shows the corresponding  
$S/N$ for measuring the power spectrum for 2 redshift bins, which is
roughly estimated as
$(S/N)_{\rm PS}\sim f_{\rm sky}^{-1/2} l_{\rm max}\sim
316 (f_{\rm sky}/0.1)^{-1/2}(l_{\rm max}/1000)$ 
(the shot noise is ignored). 
It is apparent that
the $S/N$ values for power spectrum and bispectrum 
become comparable at $l_{\rm max}\sim 1000$. Further, 
the bispectrum
$S/N$  displays stronger dependence on $l_{\rm max}$
(roughly $S/N\propto l_{\rm max}^2$ around $l_{\rm max}\sim 1000$)
than seen in power spectrum. Because
the bispectrum measurement is contributed from $l_{\rm
max}^3$ triangles, while the power spectrum from different $l_{\rm max}$
modes.

\section{Fisher matrix analysis for lensing tomography}
\label{fisher}

We will use the Fisher matrix formalism (see Tegmark et
al. 1997 for a review), to examine how tomography with
the lensing power spectrum and bispectrum can constrain
cosmological parameters and the equation of state of dark energy.

\subsection{Methodology}

Given a data vector $\bm{x}$, the Fisher
information matrix describes how the errors propagate
into the precision on parameters $p_\alpha$.  The Fisher matrix
is given by
\begin{equation}
F_{\alpha\beta}=-\kaco{\frac{\partial^2 \ln L}
{\partial p_\alpha\partial p_\beta}},
\end{equation}
where $L$ is the likelihood function of the data set $\bm{x}$
given the true parameters
$p_1,\cdots, p_\alpha$. The partial derivative with respect to a
parameter $p_\alpha$ is evaluated around the fiducial model. 
The Fisher matrix quantifies the best statistical errors achievable 
on parameter determination with a given data set:
The variance of an unbiased estimator of a parameter $p_\alpha$ obeys
the inequality: 
\begin{equation}
\skaco{\Delta p_\alpha^2}\ge (\bm{F}^{-1})_{\alpha\alpha},
\end{equation}
where $(\bm{F}^{-1})$ denotes the inverse of the Fisher matrix and
$\Delta p_\alpha$ is the relative error on parameter $p_\alpha$ around
its fiducial value. Note that this condition 
includes marginalization over the other parameters $p_\beta$
($\alpha\ne \beta$).

First, we consider power spectrum tomography. Assuming the
likelihood function for the lensing power spectrum to be Gaussian, the
Fisher matrix can be expressed as
\begin{eqnarray}
F^{\rm ps}_{\alpha \beta}=\sum_{l=l_{\rm min}}^{l_{\rm max}} \ 
\sum_{(i,j),(m,n)}
\frac{\partial C_{(ij)}(l)}{\partial p_\alpha}
\left[{\rm Cov}[C_{(ij)}(l),C_{(mn)}(l)]\right]^{-1}
\frac{\partial C_{(mn)}(l)}{\partial p_\beta},
\end{eqnarray}
where the power spectrum covariance is given by equation
(\ref{eqn:covps}) and $[{\rm Cov}]^{-1}$ denotes the inverse matrix. In
the Fisher matrix evaluation above, we consider $n_s(n_s+1)/2$ cross-
and auto-power spectra taken from $n_s$ redshift bins.
We will use the power spectrum in $100$ bins in $l$ rather
than at every multipole moment. We checked that the binning makes no
large difference on our results, since the lensing power spectrum does
not have any complex features.

Similarly, we assume that the likelihood function for the lensing
bispectrum is close to Gaussian.
The Fisher matrix for bispectrum tomography is then given by
\begin{eqnarray}
F^{\rm bisp}_{\alpha\beta}=
\sum_{l_{\rm min}\le 
l_1\le l_2\le l_3\le l_{\rm max}} \ \sum_{(i,j,k),(a,b,c)}
\frac{\partial B_{l_1l_2l_3(ijk)}}{\partial p_\alpha}
\left[{\rm Cov}[B_{l_1l_2l_3(ijk)},
B_{l_1l_2l_3(abc)}
]\right]^{-1}
\frac{\partial B_{l_1l_2l_3(abc)}}{\partial p_\beta},
\end{eqnarray}
where the bispectrum covariance is given by equation (\ref{eqn:covbisp})
and we have imposed the condition $l_1\le l_2\le l_3$ so that every
triangle configuration is counted once. 
There are $\sim 10^9$ triangles for
$l_{\rm max}=3000$, hence it is computationally intractable to
compute  the contributions from every triangle.
We bin $l_1$ and $l_2$, and use the approximation:
\begin{eqnarray}
F^{\rm bisp}_{\alpha\beta}=
\sum_{l_1'}\sum_{l_2'}
\Delta l_1\Delta l_2
\sum_{l_3} \ 
\sum_{(i,j,k),(a,b,c)}
\frac{\partial B_{l_1l_2l_3(ijk)}}{\partial p_\alpha}
\left[{\rm Cov}[B_{l_1l_2l_3(ijk)},
B_{l_1l_2l_3(abc)}
]\right]^{-1}
\frac{\partial B_{l_1l_2l_3(abc)}}{\partial p_\beta},
\label{eqn:fisherbin}
\end{eqnarray}
where $l_1', l_2'$ denote the binned values of $l_1$ and $l_2$
(we will use $100$ bins in
the following results).  Note that we continue to use a direct summation
over $l_3$ so that we can correctly account for the condition that the
Wigner 3-$j$ symbol is non-vanishing for $l_1+l_2+l_3={\rm even}$ and
vanishes for $l_1+l_2+l_3={\rm odd}$, where $l_1$ and $l_2$ are now 
the central values in the bin. We checked that our results are not
sensitive to this binning, because the lensing bispectrum is smooth 
within the bin widths used.  

We next consider the Fisher information matrix for a joint measurement
of the lensing power spectrum and bispectrum.  To do this requires
a knowledge of the cross covariance between measurements of 
the power spectrum and bispectrum, which quantifies how the two 
observables are correlated.
However, the covariance has no Gaussian contribution expressed in terms
of products of the power spectrum, and arises from the 5-point function
of the lensing fields.  Hence, we ignore the cross covariance for the
consistency of the procedure we have taken: the Fisher matrix for the
joint measurement can then be approximated by
\begin{equation}
F_{\alpha\beta}\approx F_{\alpha\beta}^{\rm ps}+F_{\alpha\beta}^{\rm bisp}. 
\end{equation}

It is useful to consider constraint ellipses in a two-parameter 
subset of the full parameter space. These are obtained by
projection of the higher dimensional ellipsoids, since the projected
ellipses include marginalization over the other parameter uncertainties.
Assuming that the form of the parameter likelihood function is
approximated by a multivariate Gaussian function, the projected
likelihood function for the two parameters $p_\alpha$ and $p_\beta$ is
given by (e.g., Matsubara \& Szalay 2002):
\begin{eqnarray}
{\cal L}(p_\alpha,p_\beta)\propto
\exp\left\{
-\frac{1}{2}
\left(
\Delta p_\alpha~ \Delta p_\beta\right)
\left[
\begin{array}{cc}
(\bm{F}^{-1})_{\alpha\alpha}&(\bm{F}^{-1})_{\alpha\beta}\\
(\bm{F}^{-1})_{\alpha\beta}&(\bm{F}^{-1})_{\beta\beta}
\end{array}
\right]^{-1}
\left(
\begin{array}{c}
\Delta p_\alpha\\
\Delta p_\beta
\end{array}
\right)
\right\},
\end{eqnarray}
where the normalization factor is determined to satisfy $\int\Delta
p_\alpha\Delta p_\beta {\cal L}=1$ and $[\cdots ]^{-1}$ denotes the
inverse matrix.  When many
parameters are considered (we will use up to 8 parameters),
the Fisher
matrix formalism has considerable computational advantages over a
grid-based search in the likelihood space.  Finally, the correlation 
coefficient, which quantifies how the constraints on
parameters $p_\alpha$ and $p_\beta$ are degenerate with each other, 
is defined as
\begin{eqnarray}
r(p_\alpha,p_\beta)=\frac{({\bm F}^{-1})_{\alpha\beta}}
{\sqrt{({\bm{F}}^{-1})_{\alpha\alpha}({\bm{F}}^{-1})_{\beta\beta}}}.
\label{eqn:coeff}
\end{eqnarray}
If $|r|=1$, the parameters constraints are totally degenerate, while
$r=0$ means they are uncorrelated. 

\subsection{The parameter space and fiducial model}

We restrict our analysis to a flat
CDM universe, supported by CMB observations (e.g,
Spergel et al. 2003). The parameter forecasts derived are sensitive to
the parameter space used and to whether constraint on a given
parameter are obtained by marginalizing 
over other parameter uncertainties. We mainly use seven 
parameters which determine the
lensing observables within the CDM model: $\Omega_{\rm de}$, $w_0$, $w_a$,
$\sigma_8$, $n$, $\Omega_{\rm b}h^2$ and $h$, where $\Omega_{\rm de}$
and $\Omega_{\rm b}$ are the density parameters of dark energy and baryons,
$n$ is the spectral index of primordial scalar perturbations, $h$ is
the Hubble parameter, and $\sigma_8$ is the rms mass fluctuation in a
sphere of radius $8h^{-1}$Mpc. We use a simple parameterization of 
the equation of state of dark
energy: $w(a)=w_0+w_a(1-a)$ (Turner \& White 1997; Linder 2003).  
Note that we use $\Omega_{\rm b}h^2$ rather than $\Omega_{\rm b}$, 
because $\Omega_{\rm b}h^2$ is
directly probed by the CMB observations and we will use its prior
constraint in the Fisher analysis.  For the input linear mass power
spectrum, we employ the BBKS transfer function
(Bardeen et al. 1986) with the shape parameter given by Sugiyama (1995),
which accounts for the baryon contribution. We thus ignore the effect of
dark energy spatial fluctuations on the transfer function.  For slowly
varying dark energy models with $w\le -0.6$ (the rough current limit from 
observations, Perlmutter, White \& Turner 1999), the BBKS
approximation is reliable on scales probed by weak lensing surveys.
Note that dark energy clustering affects horizon scales, and
it therefore modifies the COBE normalization through changes to the
Sachs-Wolfe effect and the integrated Sachs-Wolfe 
effect (Caldwell et al. 1998; Ma
et al. 1999; Hu 2002a). 
 For precise comparisons with data, we will need
more accurate transfer function as given by CMBFAST (Seljak \&
Zaldarriaga 1996).  Finally, in this paper we ignore the possible effect of 
massive neutrinos on the mass power spectrum (see e.g. Hu 2002a; 
Abazajian \& Dodelson 2003).

The Fisher matrix formalism assesses how well lensing observables
can distinguish the true (``fiducial'') model of the universe from 
other models. The results we obtain depend upon the fiducial model. 
We use the currently concordant flat $\Lambda$CDM model
with $\Omega_{\rm de}=0.65$, $w_0=-1$, $w_a=0$, $\sigma_8=0.9$, $n=1$,
$\Omega_{\rm b}=0.05$ and $h=0.72$, which is consistent with the recent
WMAP results (Spergel et al. 2003).  To compute the Fisher matrix, we
need to choose steps in parameter directions to compute the numerical
derivatives. We choose the steps to be $5\%$ in the parameter values, except
for $\Delta w_a=\pm 0.1$; two-side numerical derivatives are used. 
 We will pay special attention to how parameters $\Omega_{\rm
de}$, $w_0$, $w_a$ and $\sigma_8$ can be constrained from lensing
tomography.  On the other hand, we will employ prior information on
$n$ $\Omega_{\rm b}h^2$ and $h$, expected from
measurements of the CMB temperature and polarization by the Planck
satellite mission (e.g., Eisenstein et al. 1999).  Assuming Gaussian
priors, we add the diagonal component $(F_{\rm
prior})_{\alpha\beta}=\delta_{\alpha\beta} \sigma(p_\alpha)^{-2}$ to the
Fisher matrix. 
Note that we do not employ any priors to the parameters $\Omega_{\rm
de}$, $w_0$, $w_a$ and $\sigma_8$ and thus the following results are
conservative (see Figure \ref{fig:lensCMB} for the joint parameter
forecasts using lensing tomography and the CMB).

\section{Results: Cosmological constraints}
\label{result}

\begin{figure*}
  \begin{center}
    \leavevmode\epsfxsize=14.cm \epsfbox{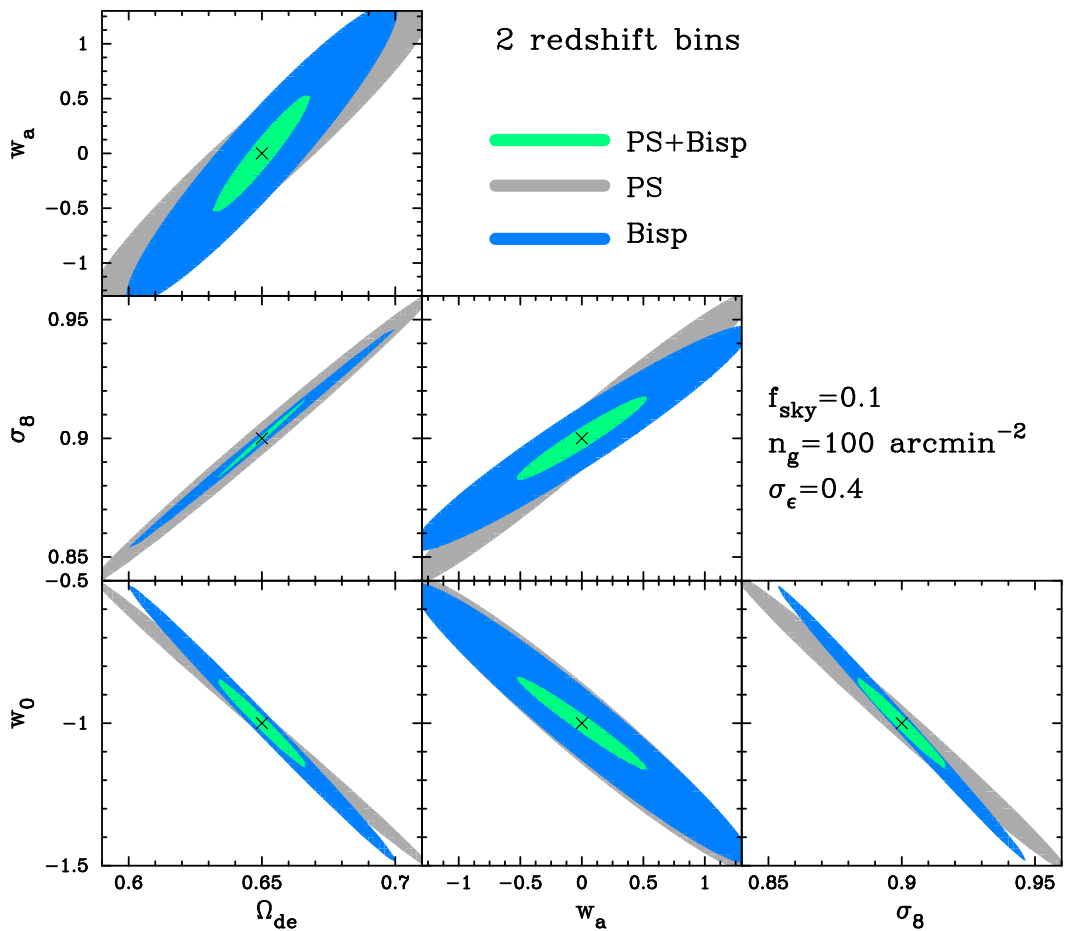}
  \end{center}
\caption{Projected $68\%$ CL constraints in the parameter space of
$\Omega_{\rm de}$, $w_0$,
$w_a$ and $\sigma_8$ from the lensing power spectrum and 
the bispectrum in two redshift bins, as indicated. We employ $7$
cosmological parameters in the Fisher matrix analysis. The results
shown are obtained assuming priors on
$n$, $\Omega_{\rm b}h^2$ and $h$
 expected from the Planck mission. The sky coverage and number density
are taken to be $f_{\rm sky}=0.1$ and $n_g=100 $ arcmin$^{-2}$,  
and angular modes $50\le l\le 3000$ are used. 
It is clear that bispectrum tomography can improve parameter constraints 
significantly, typically by a factor of three, compared to just power
spectrum tomography. 
}
\label{fig:chins2}
\end{figure*}

We consider three scenarios for constraining cosmological parameters: 
using the lensing power spectrum and bispectrum without any
tomography, tomography in two redshift bins, and in three redshift
bins. 
Figure \ref{fig:chins2} shows the resulting
constraint ellipses in the parameter space of $\Omega_{\rm de}$, $w_0$, $w_a$
and $\sigma_8$, corresponding to the $68\%$ confidence level ($\Delta
\chi^2=2.3$), marginalized over the parameters $\Omega_{\rm
b}$, $n$ and $h$.  
We consider angular scales
$50\le l\le 3000$ and assume the priors $\sigma(\ln
\Omega_{\rm b}h^2)=0.010 $, $\sigma(n)=0.008$ and $\sigma(h)=0.13$,
expected from the Planck mission (see Table 2 in Eisenstein et al. 1999).  
\begin{figure*}
  \begin{center}
    \leavevmode\epsfxsize=14.cm \epsfbox{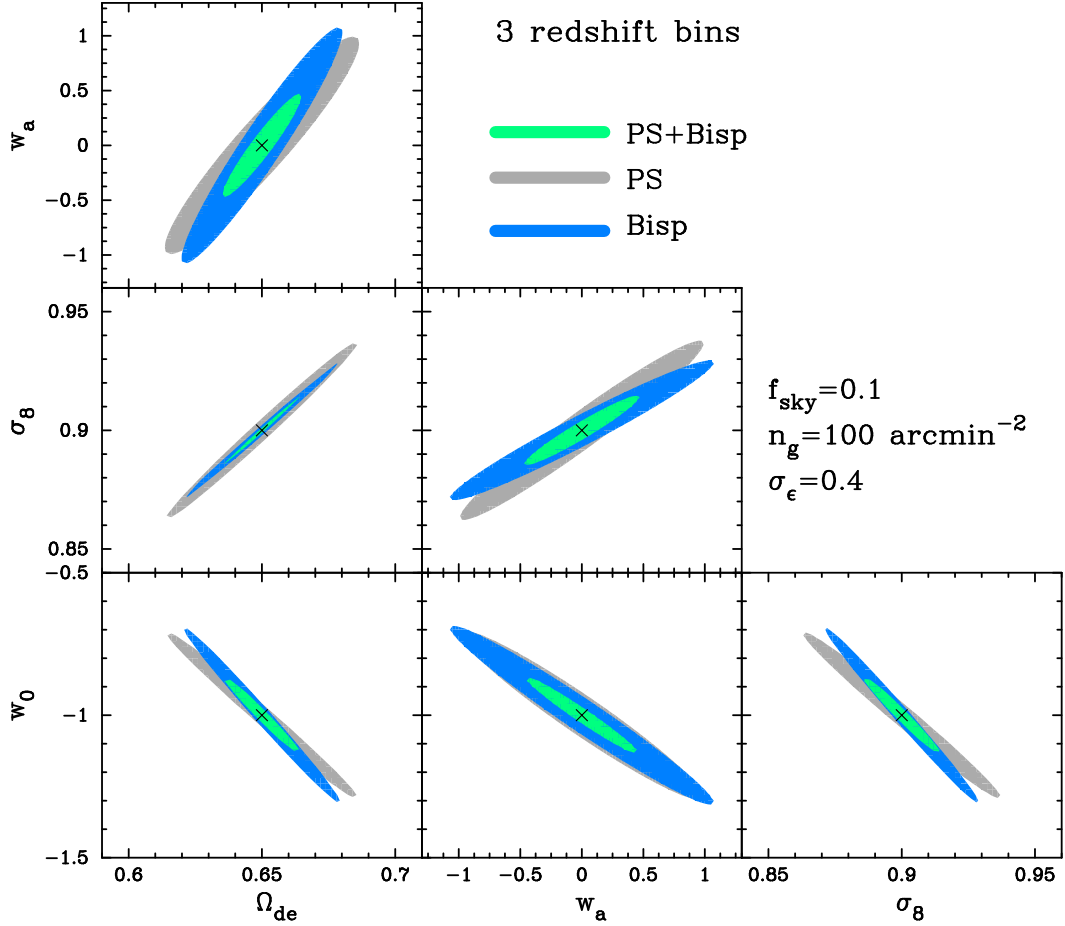}
  \end{center}
\caption{As in the previous figure, but for 3 redshift
 bins. In this case, we use 6 power spectra and 27 bispectra in the
 analysis.  
}
\label{fig:chins3}
\end{figure*}
Each redshift bin is chosen so as to have equal
number density of galaxies; $0\le z\le 1.3 $ and $1.3\le z$ for the
galaxy distribution shown in Figure \ref{fig:pz}.  
One can see that
bispectrum tomography provides additional information on the parameters
relative to power spectrum tomography.  
In particular, one striking result
is that the parameter constraints from bispectrum tomography
alone are comparable with those from power spectrum tomography.
As a result, combining the two can significantly improve 
parameter determination by a factor of $3$, because of their complementarity.
For example, the equation of state parameters
are constrained as $\sigma(w)=0.034f_{\rm sky}^{-1/2}$ and
$\sigma(w_a)=0.11f_{\rm sky}^{-1/2}$.
There are two essential features needed 
to obtain these results.  First, we considered all triangle
configurations ($\sim 10^9$
triangles) available from a range of angular scales we consider, which
leads to a significant gain in $S/N$. Second, we considered all
the cross- and auto-spectra constructed from the redshift bins,
providing $3$ power spectrum and $8$ bispectrum contributions to 
the Fisher matrix. 

The correlation coefficients of the Fisher matrix elements quantify how
the parameter constraints are degenerate.  From the
definition (\ref{eqn:coeff}), we find the dark energy parameters are
highly correlated as $r(\Omega_{\rm de},w_0)=-0.98$, $r(\Omega_{\rm
de},w_a)=0.95$ and $r(w_0,w_a)=-0.97$.  It is also worth noting the
importance of the marginalization in the Fisher matrix formalism. If we
fix $\Omega_{\rm b}$, $n$ and $h$ to their fiducial values, the
constraints improve significantly: $\sigma(\Omega_{\rm de})=0.0031$,
$\sigma(w_0)=0.038$, $\sigma(w_a)=0.15$ and $\sigma(\sigma_8)=0.0035$ 
compared to the errors shown in Table \ref{tab:const}.  

Table 1 gives the $1$-$\sigma$ errors on parameters for 
three cases: no tomography, tomography with 2 redshift bins, and
with 3 redshift bins. 
Clearly tomography leads to significant improvements in
parameter accuracies. For the equation of state parameters
$w_0$ and $w_a$, the improvement in the 1-$\sigma$ errors 
is a factor of $10$.  
As stressed in Hu (1999), on the other hand, 
fine divisions of the galaxy redshift
distribution for lensing tomography does not give much additional
information. 
This can be seen from Figure \ref{fig:chins3}, 
which shows the results for three redshift bins, $0\le z\le 1$, $1\le
z\le 1.7$ and $z\ge 1.7$: 
In this case, we considered 27 bispectra and 6 power spectra in the Fisher
matrix analysis. The constraints on the parameters are 
improved by $15$-20$\%$.
\begin{table}
\begin{center}
Parameter Estimation\\
\begin{tabular}{l\colskip ccc\colskip ccc\colskip ccc}\hline\hline
&\multicolumn{3}{c}{\hspace{-2em}no tomography} 
&\multicolumn{3}{c}{\hspace{-2em} 2 redshift bins}
&\multicolumn{3}{c}{3 redshift bins}\\
Quantity & PS & Bisp &PS+Bisp & PS & Bisp &PS+Bisp &PS&Bisp &PS+Bisp\\ \hline
$\sigma(\Omega_{\rm de})$& 0.14& 0.13& 0.023  
&0.043&0.034&0.012 
&0.024&0.020&0.0098 (0.0046)\\
$\sigma(\sigma_8)$&0.12&0.12&0.023  
&0.041&0.031&0.012 
&0.025&0.020&0.0098\\
$\sigma(w_0)$&4.0&3.6&1.1  
&0.34&0.32&0.11 
&0.20&0.21&0.092 (0.046)\\
$\sigma(w_a)$&20&15&5.1 
&0.93&0.91&0.36 
&0.65&0.71&0.31 (0.14) \\ \hline
\end{tabular}
\end{center}
\caption{Summary of parameter constraints from lensing 
 tomography of the power spectrum (PS), the bispectrum (Bisp) and both
 measurements (PS+Bisp).  
All errors are
 $1$-$\sigma$ level and include marginalization over the other
 parameters. 
Note that we have used $f_{\rm sky}=0.1$ 
 and all the $1\sigma$ errors scale as $\propto f_{\rm
 sky}^{-1/2}$.
The values in the brackets for three
redshift bins show the constraints when we combine lensing
 tomography and CMB observations  (see Figure \ref{fig:lensCMB} in more detail). 
}  \label{tab:const}
\end{table}

\begin{figure*}
  \begin{center}
    \leavevmode\epsfxsize=17.cm \epsfbox{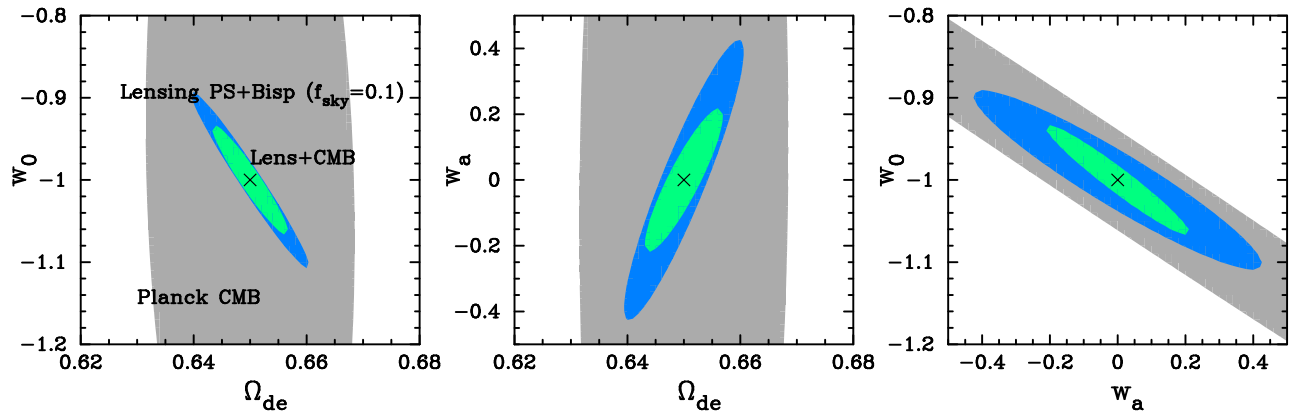}
  \end{center}
\caption{Error ellipses ($68\%$ CL) for dark energy parameters
($\Omega_{\rm de},w_0,w_a$) when we use lensing tomography
(PS+Bisp) with three redshift bins, 
CMB power spectra from the Planck experiment
and combine the two. 
The constraints include 
 marginalization over the other parameters, including the optical depth (see
 the text in more detail). 
}
\label{fig:lensCMB}
\end{figure*}
\begin{figure*}
  \begin{center}
    \leavevmode\epsfxsize=17.cm \epsfbox{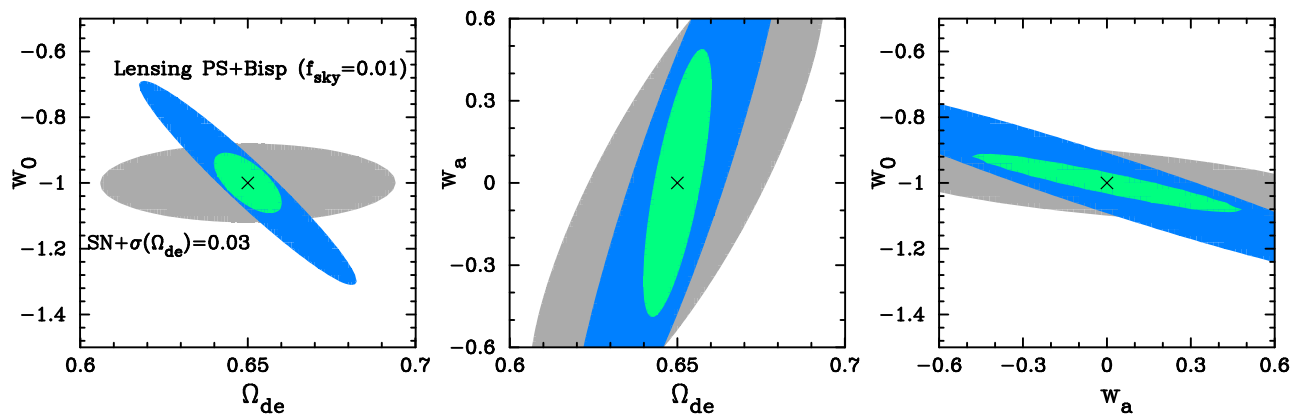}
  \end{center}
\caption{Parameter accuracy forecasts for the proposed SNAP survey of 
Type Ia supernovae and weak lensing, and the joint constraints from
the two measuremetns.  Note that the sky coverage $f_{\rm sky}=0.01$ is
assumed for the lensing survey and we used the prior
$\sigma(\Omega_{\rm de})=0.03$ for the supernova constraint. 
} \label{fig:chiovwwa}
\end{figure*}
As shown in Figures \ref{fig:chins2} and \ref{fig:chins3}, lensing
tomography alone leads to degenerate constraints on $\Omega_{\rm
de}$ and the equation of state parameters $w_0$ and $w_a$.  Therefore,
external information on these parameters from other methods such as
observations of CMB, supernovae and galaxy redshift surveys is valuable
in that it allows us to break the degeneracies. This is investigated in
Figure \ref{fig:lensCMB}, which shows how the constraint ellipses in the
subspace of ($\Omega_{\rm de}$, $w_0$, $w_a$) change when we combine the
constraints from lensing tomography and CMB.  To compute the Fisher
matrix from measurements of the CMB temperature and polarization power
spectra and the cross spectrum, we assumed the experimental
specifications of the Planck 143 and 217\ GHz channels with $65\%$ sky
coverage and assumed a spatially flat
universe, no massive neutrinos, no running tilt and no gravity
waves. However, we include the dependence of the dark energy equation of
state, which modifies the CMB power spectra through the change of the
angular diameter distance to the last scattering surface (we ignored its
spatial fluctuations). 

In this case, the CMB power spectra are specified
by 8 parameters ($\Omega_{\rm de}$, $w_0$, $w_a$, $\delta_\zeta$, $n$,
$\Omega_{\rm b}h^2$, $h$, $\tau$) and have been computed using  
CMBFAST version 4.5 (Seljak \& Zaldarriaga 1996).  $\delta_\zeta$ is the
amplitude of the primordial curvature fluctuation and $\tau$ denotes the
optical depth (their fiducial values are taken to be $\delta_\zeta
=4.56\times 10^{-5}$ and $\tau=0.15$). Likewise, to compute the Fisher
matrix from lensing tomography, we employed the $\delta_\zeta$
normalization of the mass power spectrum instead of the $\sigma_8$
normalization for the consistency.  Figure \ref{fig:lensCMB} shows that
combining the constraints from CMB measurements and lensing tomography
significantly improves the dark energy constraints, yielding
$\sigma(\Omega_{\rm de})=4.6\times 10^{-3}$, $\sigma(w_0)=0.046$ and
$\sigma(w_a)=0.14$. Note that the results shown are derived by adding
the two $(7\times 7)$ Fisher matrices from lensing tomography and the
CMB spectra, including the marginalization over the optical depth. 
It is also worth noting that the CMB measurements can
better constrain the other parameters $(\delta_\zeta,\Omega_b,n,h)$ than
lensing tomography, which is part of the reason for the improvement
in Figure \ref{fig:lensCMB}.

 Figure \ref{fig:chiovwwa} shows forecasts of the dark
energy constraints expected from the SNAP survey of 
 Type Ia supernova measurement (E. Linder, private
communication) and its lensing survey.  We assumed the sky coverage $f_{\rm
sky}=0.01$ for the SNAP weak lensing survey (see Massey et al. 2003;
Refregier et al. 2003).  It is again apparent that a joint analysis of 
weak lensing and supernovae significantly improves the parameter errors
due to their complementarity: the errors become $\sigma(\Omega_{\rm
de})=6.7\times 10^{-3}$, $\sigma(w_0)=0.061$ and $\sigma(w_a)=0.32$.
Moreover, this result can be further improved by combining with the CMB
constraints, which gives (for $f_{\rm sky}=0.01$): 
$\sigma(w_0)=0.042$ and $\sigma(w_a)=0.15$.

\subsection{Constraints on running spectral index}

Next we examine how lensing observations can probe the
running spectral index of primordial scalar perturbations, 
motivated from the WMAP result (Spergel et
al. 2003).  CMB measurements probe large length scale
fluctuations and the constraint on the running spectral index is to some
extent diluted by the other parameter uncertainties (especially the
optical depth $\tau$).  In fact, strong constraints on the running
of the spectral index can be obtained only by combining
CMB, galaxy surveys and the Ly-$\alpha$ forest. 
Weak lensing directly probes the relevant
small length scales, and is not susceptible to uncertainties due to 
biasing. The measured spectra however have undergone nonlinear
evolution, so an accurate model is needed to infer the shape of the
linear spectrum. 

\begin{figure*}
  \begin{center}
    \leavevmode\epsfxsize=13.cm \epsfbox{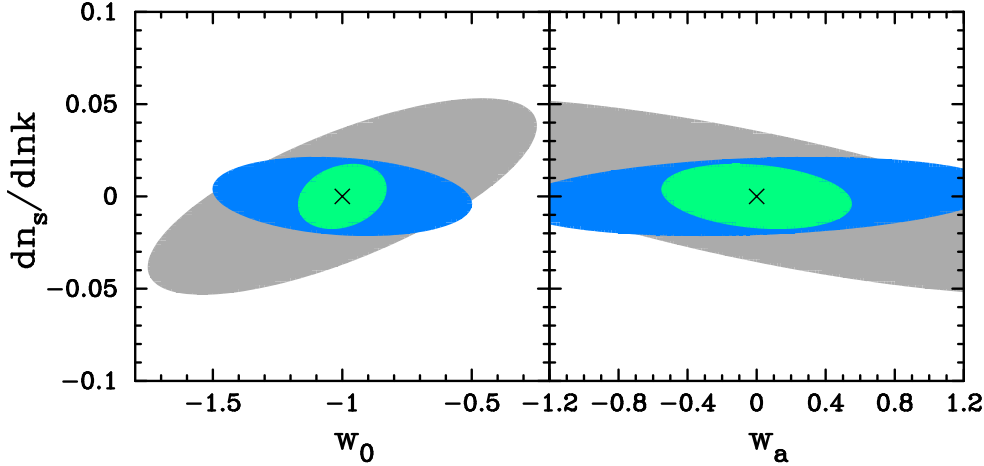}
  \end{center}
\caption{The constraint ellipses in the parameter space of
 $w_0$, $w_a$ and the running spectral index parameter 
$dn/d\ln k$. The Fisher analysis is as for Figure
 \ref{fig:chins2}. In total, 8 parameters are considered 
 to obtain the marginalized ellipses shown ($dn/d\ln k$, 
 plus the seven parameters considered in the preceding analysis). 
}
\label{fig:chidns}
\end{figure*}
To make forecasts for future lensing surveys, we
follow the model in Spergel et al. (2003), where the linear mass power
spectrum is expressed as
\begin{equation}
P_\delta(k)\propto \left(\frac{k}{k_0}\right)^{n + 
\alpha_s \ln(k/k_0)/2}T^2(k), 
\label{eqn:runns}
\end{equation}
where $\alpha_s=dn/d\ln k$ and 
$k_0=0.05 $ Mpc$^{-1}$ and $T(k)$ is the transfer function. 
Figure \ref{fig:chidns} shows the forecast
constraints on $w_0$, $w_a$ and $dn/d\ln k$, for the two redshift
bins. Here we considered 8 parameters in the Fisher matrix
analysis ($dn/d\ln k$, plus the seven parameters we have used so
far), which enlarges the constraint ellipses in Figure \ref{fig:chins2}.
It is clear that lensing tomography can put stringent constraints on the
running spectral index: $\sigma(dn/d\ln k)=0.0037f_{\rm
sky}^{-1/2}$. This result can be compared
with the constraint $\sigma(dn/d\ln k)=0.004$ expected from the Planck
mission (see Table 2 in Eisenstein et al. 1999) and with the constraints
expected from galaxy redshift surveys ($\sigma(dn/d\ln k)=0.42$)
and the Ly-$\alpha$ forest ($\sigma(dn/d\ln k)=0.04$ in Mandelbaum et
al. 2003) \footnote{Note that the definition of $k_0$ in the model
(\ref{eqn:runns}) differs in these papers, which affects the
$1$-$\sigma$ error.}. For this parameter the bispectrum is especially
powerful, as evident from Figure \ref{fig:chidns}. 

\section{Conclusion and Discussion}
\label{conc} 

We have used lensing tomography with
the power spectrum and bispectrum as a
probe of dark energy evolution and the mass power spectrum. 
The lensing bispectrum has
different dependences on the lensing weight function and the
growth rate of mass clustering from those of the power spectrum. 
Thus bispectrum tomography provides complementary
constraints on cosmological parameters to power spectrum tomography. 
By using information from different triangle configurations and
all cross-spectra in redshift bins we find that the bispectrum has
roughly as much information as the power spectrum on parameters of interest
(see Figures \ref{fig:chins2} and \ref{fig:chins3}).  
Parameter accuracies are typically
improved by a factor of $3$ if both the power spectrum and bispectrum 
are used, compared to the standard approach of using just the power
spectrum (see Table 1). 
Thus our study provides strong motivation for the use of
the bispectrum from lensing surveys for parameter extraction. 

Since lensing observables are significantly affected by non-linear
gravitational clustering on angular scales below a degree,  we
used a non-linear model (see discussion below) to compute the lensing 
bispectrum and to estimate the precision on
the parameters $\sigma_8$, $\Omega_{\rm de}$, $w_0$, $w_a$ and 
$dn/d\ln k$. 
The constraints on the dark energy parameters are $\sigma(w_0)\sim 0.03f_{\rm
sky}^{-1/2}$ and $\sigma(w_a)\sim 0.1 f_{\rm sky}^{-1/2}$ -- this
sensitivity to the redshift evolution of the equation of state of 
dark energy is comparable to the best methods proposed for the coming
decade. 
Moreover, external information on dark energy parameters such as provided
by CMB and Type Ia supernova measurements 
can significantly improve the parameter accuracies as shown in 
Figures \ref{fig:lensCMB} and \ref{fig:chiovwwa}.
In addition, lensing tomography can precisely probe the
mass power spectrum: the constraint on the power spectrum amplitude is 
$\sigma(\sigma_8) \sim 4\times 
10^{-3}f_{\rm sky}^{-1/2}$, and on the running spectral 
$\sigma(dn/d\ln k) \sim  4\times 10^{-3}f_{\rm sky}^{-1/2}$. 
Our analysis includes the full information at the two- and three-point
level. Using three-point statistics such as the skewness, which
contain no information on triangle configurations, 
weakens parameter constraints significantly. 

Thus there is strong motivation to build an accurate 
model of lensing observables in the moderately non-linear regime. 
This should be feasible, since the physics
involved in lensing is only gravity. It will also be
necessary to calibrate the covariance of
the bispectrum over relevant triangle configurations using a sufficient
number of simulation realizations. 
Such an accurate model of the lensing bispectrum might modify 
the ellipse shapes in Figure \ref{fig:chins2}.
Even so, we believe that the level of
improvement from bispectrum tomography
is likely to be correct, because the mass
bispectrum we have employed should correctly estimate the
amplitude, and therefore the signal-to-noise to within $10$-$20\%$. 

We also estimate how bispectrum tomography can put constraints on 
primordial non-Gaussianity.  The most optimal constraint was
estimated by requiring 
that the bispectrum due to primordial non-Gaussianity is
detectable with $S/N\ge 1$. We neglected the bispectrum induced
by gravitational non-Gaussianity, since it differs
in its redshift evolution and configuration
dependence. However, even with this assumption, the result
in Figure \ref{fig:fnl} is not very promising: the constraint on
the primordial non-Gaussian model
is not as stringent as that from the CMB
bispectrum measurement (Komatsu et al. 2003), 
unless an almost all-sky lensing survey is available. 
It is worth exploring how the three-dimensional mass reconstruction
proposed by Hu \& Keeton (2003) could allow us to improve 
the lensing estimates. 

We have concentrated on statistical measures of the convergence. 
However, the convergence field is not a direct observable, and
reconstructing the convergence field from the measured ellipticities
(shear) of galaxies is still challenging from survey data. 
Reconstruction techniques have been proposed for 
the power spectrum and the convergence field from 
realistic data (see Kaiser 1998; Hu \& White 2001 for the 2D case and
Taylor 2003 and Hu \& Keeton 2003 for 3D mass reconstruction). 
It is of interest to develop an optimal method of extracting 
the lensing bispectrum from the measured shear field. Alternatively, we can use
measurements of the three-point correlation functions of the shear 
fields, which have been extensively studied recently by Schneider \&
Lombardi (2003), Zaldarriaga \& Scoccimarro (2003) and TJ03a,c
(see Bernardeau, Mellier \& Van Waerbeke 2003; 
Pen et al. 2003; Jarvis, Bernstein \& Jain 2003 for measurements).
The shear three-point correlation functions
carry full information on the convergence bispectrum 
(Schneider,  Kilbinger \& Lombardi 2003).
Therefore, all the results derived in
this paper are attainable using tomography of the two- and
three-point functions of the shear fields.

There are other uncertainties we have ignored in this paper. We have
assumed accurate photometric redshift measurements.
For finite errors in the redshifts, we cannot take narrow
redshift subdivisions of the galaxy distribution. 
This error would lead to additional statistical errors on measurements
of the power spectrum and bispectrum and in turn 
on the cosmological parameters. Since we use only two or three redshift
bins in our analysis, the demands on statistical errors are not very
stringent. However, possible biases
in the photometric redshifts must be carefully examined, because the redshift
evolution of dark energy leads to only a small effect on the lensing
observables, as discussed in Bernstein \& Jain (2003).  

We have also ignored the $B$-mode contamination to the lensing
observables, though it is seen in current measurements
(e.g., Jarvis et al. 2003). The main source is likely to be
observational systematics which are not eliminated in the PSF
correction. For a shallow survey, the intrinsic ellipticity alignments
could also provide significant contribution to the $B$-mode on small 
angular scales. Another useful
application of photometric redshift information is that it allows us
to remove intrinsic alignment contaminations by excluding close pairs of
galaxies in the same redshift bin (Heymans \& Heavens 2003; King
\& Schneider 2003). Thus the cross-power spectrum calculated
from different redshift bins is not affected by the intrinsic
alignment. 
Hence, lensing tomography that uses the
cross-power spectra and  bispectra is robust
to systematics (Takada \& White 2003). 
The degradation in parameter accuracies is likely to be
small, because the cross-power spectrum carries comparable 
lensing signal to the auto-spectrum (see Figure \ref{fig:cl} and Takada
\& White 2003).

\bigskip

We thank G. Bernstein, D. Dolney, M. Jarvis, 
E. Komatsu, W. Hu, L. King, E. Linder, 
S. Majumdar, A. Refregier, P. Schneider, 
R. Scoccimarro and M. White for useful discussions. 
We also thank U. Seljak
and M. Zaldarriaga for making updated versions of 
CMBFAST publicly available. 
We acknowledge the hospitality of the Aspen Center for Physics, 
where this work was begun.  This work is supported by NASA grants
NAG5-10923, NAG5-10924 and a Keck foundation grant.

\appendix
\section{Approximation for Wigner-3$j$ Evaluation}

The expression for the bispectrum of the convergence field involves the
Wigner-3$j$ symbol, which has a closed algebraic form (e.g., Hu 2000):
\begin{equation}
\left(
\begin{array}{ccc}
l_1&l_2&l_3\\
0&0&0
\end{array}
\right)=(-1)^{L/2}\frac{(\frac{L}{2})!}{\left(\frac{L}{2}-l_1\right)!
\left(\frac{L}{2}-l_2\right)!\left(\frac{L}{2}-l_3\right)!}
\left[\frac{(L-2l_1)!(L-2l_2)!(L-2l_3)!}{(L+1)!}\right]^{1/2},
\label{eqn:w3j}
\end{equation}
for even $L=l_1+l_2+l_3$ and zero for odd $L$. 

Rather than using the exact equation, we employ an approximation of the
Wigner-3$j$ symbol evaluation, since the direct numerical calculation
encounters a divergence problem in the factorial as $l!$ for large
$l$. For this purpose, we use the Stirling approximation:
$n!=\Gamma(n+1)$ and 
\begin{equation}
\Gamma(x)\sim (2\pi)^{1/2}e^{-x}x^{x-1/2},~ ~ \mbox{  for large  }x. 
\end{equation}
This approximation is quite good, leading to errors less than $0.2\%$
for angular scales $l\ge 50$ of our interest. Using this approximation,
the expression (\ref{eqn:w3j}) of the Wigner-3$j$ symbol is rewritten
as
\begin{eqnarray}
\left(
\begin{array}{ccc}
l_1&l_2&l_3\\
0&0&0
\end{array}
\right)
&\approx& (-1)^{L/2}(2\pi)^{-1/2}e^{3/2}2^{1/4}
\left(L+2\right)^{-1/4}
\left(\frac{\frac{L}{2}-l_1+\frac{1}{2}}
{\frac{L}{2}-l_1+1}\right)^{\frac{L}{2}-l_1+\frac{1}{4}}
\left(\frac{\frac{L}{2}-l_1+\frac{1}{2}}
{\frac{L}{2}-l_1+1}\right)^{\frac{L}{2}-l_2+\frac{1}{4}}
\nonumber\\
&&\times \left(\frac{\frac{L}{2}-l_1+\frac{1}{2}}
{\frac{L}{2}-l_1+1}\right)^{\frac{L}{2}-l_3+\frac{1}{4}}
\frac{1}{
\left(\frac{L}{2}-l_1+1\right)^{1/4}
\left(\frac{L}{2}-l_2+1\right)^{1/4}
\left(\frac{L}{2}-l_3+1\right)^{1/4}
}.
\end{eqnarray}

\label{lastpage}
\end{document}